\documentclass[twocolumn]{aastex701}
\usepackage{amsmath, amsthm, amssymb}
\usepackage{mathrsfs}
\usepackage{bm} 
\usepackage[ruled,vlined]{algorithm2e}
\SetKwInput{KwIn}{Input}
\SetKwInput{KwOut}{Output}
\SetCommentSty{texttt} % 可选：让注释用等宽字体，和代码区分开
\usepackage{color}
\usepackage{listings}
\usepackage{xcolor}
\usepackage{tikz}

\begin{document}

\title{Foreground Subtraction with a Tensor-Based Oriented Singular Value Decomposition Method for HI Experiments}

\author[orcid=0000-0003-3858-6361,sname='Zuo']{Shifan Zuo}
%\affiliation{National Astronomical Observatories, Chinese Academy of Sciences, Beijing 100101, China}
%\affiliation{State Key Laboratory of Radio Astronomy and Technology, Beijing 100101, China}
\affiliation{State Key Laboratory of Radio Astronomy and Technology, National Astronomical Observatories, CAS, Beijing 100101, China}
\affiliation{School of Astronomy and Space Science, University of Chinese Academy of Sciences, Beijing 100049, China}
\email[show]{sfzuo@bao.ac.cn}

\author[orcid=0000-0001-6475-8863,sname='Chen']{Xuelei Chen}
%\affiliation{National Astronomical Observatories, Chinese Academy of Sciences, Beijing 100101, China}
%\affiliation{School of Astronomy and Space Science, University of Chinese Academy of Sciences, Beijing 100049, China}
%\affiliation{State Key Laboratory of Radio Astronomy and Technology, Beijing 100101, China}
\affiliation{State Key Laboratory of Radio Astronomy and Technology, National Astronomical Observatories, CAS, Beijing 100101, China}
\affiliation{School of Astronomy and Space Science, University of Chinese Academy of Sciences, Beijing 100049, China}
\affiliation{Key Laboratory of Cosmology and Astrophysics (Liaoning) \& College of Sciences, Northeastern University, Shenyang 110819, China}
\email[show]{xuelei@cosmology.bao.ac.cn}  

\author[orcid=0000-0002-1301-3893,sname='Mao']{Yi Mao}
\affiliation{Department of Astronomy, Tsinghua University, Beijing 100084, China}
\email[show]{ymao@tsinghua.edu.cn}  

\begin{abstract}
We introduce a native tensor-based framework for foreground mitigation in 21\,cm intensity mapping (IM), utilizing the Oriented Singular Value Decomposition (O-SVD) algorithm. While 21\,cm IM is a powerful probe of the large-scale structure of the Universe, its efficacy is severely limited by astrophysical foregrounds that are orders of magnitude brighter than the cosmological signal. Traditional mitigation strategies often necessitate flattening multidimensional data cubes into two-dimensional matrices, a process that potentially compromises the intrinsic spatial-spectral correlations by treating distinct spatial pixels as independent samples. By treating multi-frequency sky maps and angular power spectra as third-order tensors, the O-SVD method performs decomposition directly on the multilinear manifold, preserving the underlying physical topology and leveraging the distinct coherence properties of astrophysical foregrounds across different dimensions. We demonstrate the performance and versatility of the O-SVD framework through its application to high-fidelity simulations from the SKA Science Data Challenge 3a (SDC3a) and real-world observational data from the Tianlai Cylinder Pathfinder Array. Our results indicate that O-SVD provides a robust and universal approach for foreground subtraction, achieving high-fidelity signal recovery while offering superior performance compared to conventional matrix-based Singular Value Decomposition (SVD) methods.
\end{abstract}

\keywords{\uat{Cosmology}{343} --- \uat{H I line emission}{690} --- \uat{Dark energy}{351} --- \uat{Radio interferometers}{1345} }

\section{Introduction} \label{sec:intro}  

The 21\,cm hyperfine transition of neutral hydrogen (HI) serves as a preeminent cosmological probe, offering a unique window into the large-scale structure of the Universe across a vast range of redshifts \citep{2024arXiv241108113P}. Unlike traditional discrete galaxy surveys, 21\,cm intensity mapping (IM) enables the continuous, three-dimensional mapping of the cosmic matter distribution by measuring the aggregate emission from unresolved HI reservoirs. This technique provides a powerful means to investigate the fundamental physics of the Cosmic Dawn (CD), the Epoch of Reionization (EoR), and the late-time post-reionization era, where the 21\,cm signal acts as a tracer for dark energy and the growth of structure \citep{2020PASP..132f2001L}.

The pursuit of the redshifted 21\,cm signal has catalyzed the development of an extensive array of low-frequency radio facilities. Early milestone detections were achieved through cross-correlation analyses using single-dish telescopes such as Parkes and the Green Bank Telescope (GBT) in conjunction with optical redshift surveys \citep{2013ApJ...763L..20M, 2018MNRAS.476.3382A}. More recently, global signal experiments like EDGES and SARAS have targeted the sky-averaged 21\,cm spectrum to probe the thermal history and ionization state of the early intergalactic medium (IGM) \citep{2018Natur.555...67B, 2022NatAs...6.1473B}. Simultaneously, numerous radio interferometers have been deployed to map the spatial fluctuations of the 21\,cm signal. At low frequencies ($z \gtrsim 6$), experiments such as GMRT \citep{2011MNRAS.413.1174P}, 21CMA \citep{2007mru..confE..17W}, PAPER \citep{2010AJ....139.1468P}, LOFAR \citep{2013A&A...556A...2V}, MWA \citep{2013PASA...30....7T}, HERA \citep{2017PASP..129d5001D}, and NenuFAR \citep{2012sf2a.conf..687Z} target the CD and EoR. At intermediate frequencies ($z \lesssim 2.5$), arrays like CHIME \citep{2014era..conf10102V}, Tianlai \citep{2012IJMPS..12..256C} focus on detecting Baryon Acoustic Oscillations (BAO) to constrain the expansion history of the Universe. The new generation of interferometers, the Square Kilometre Array (SKA) \citep{2015aska.confE...1K} also includes 21\,cm cosmology as one of its key science drivers.

Despite its scientific potential, the detection of the 21\,cm signal is hindered by astrophysical foregrounds—primarily Galactic synchrotron emission and extragalactic point sources—which exceed the cosmological signal by four to five orders of magnitude \citep{2017MNRAS.464.4995M, 2018AJ....156...32E}. A fundamental distinction exploited for foreground mitigation is the spectral contrast: while foregrounds are characterized by smooth, power-law spectra, the 21\,cm signal exhibits significant spectral structure due to the evolving distribution of HI along the line of sight.

Contemporary foreground mitigation strategies generally fall into three categories: subtraction, avoidance, and suppression \citep{2019arXiv190912369C}. Subtraction techniques typically rely on parametric modeling, such as low-order polynomial fitting \citep{2006ApJ...650..529W} and Gaussian Process Regression (GPR) \citep{2018MNRAS.478.3640M}, or non-parametric blind source separation (BSS) methods like Principal Component Analysis (PCA) \citep{2013ApJ...763L..20M, 2015MNRAS.447..400A}, Independent Component Analysis (ICA) \citep{2012MNRAS.423.2518C}, Generalized Morphological Component Analysis (GMCA) \citep{2020MNRAS.499..304C}, Generalized Needlet Internal Linear Combination (GNILC) \citep{2016MNRAS.456.2749O}. Semiblind method like Singular Vector Projection (SVP) \citep{2023ApJ...945...38Z} method has also been developed. While effective, these methods often treat the data as isolated 1D spectral ``fibers'' or require flattening the 3D data cube into a 2D matrix, a process that potentially discards or distorts the inherent spatial-spectral topology essential for robust signal separation.

In this work, we propose a native tensor-based framework for 21\,cm foreground subtraction. Many data representations in 21\,cm cosmology, including multi-frequency image cubes, multi-frequency angular power spectra (MAPS), and 3D power spectra, possess an intrinsic multilinear structure that is naturally captured by tensors. The Oriented Singular Value Decomposition (O-SVD; \citealt{Zeng2020})—a hierarchical multilinear generalization of the matrix SVD—was originally introduced in the context of multilinear algebra. In this work, we apply O-SVD to perform foreground subtraction directly on the tensor manifold. This approach preserves the multi-dimensional correlations and provides a more flexible set of degrees of freedom for isolating foreground modes. We demonstrate the efficacy of the O-SVD framework using both simulated data from the SKA Science Data Challenge 3a (SDC3a) and observational data from the Tianlai Cylinder Pathfinder Array. To our knowledge, this work represents the first application of O-SVD to 21\,cm foreground mitigation, which constitutes the novel contribution of this paper.

The remainder of this paper is organized as follows: Section~\ref{sec:method} details the mathematical foundation of the O-SVD framework and its implementation for foreground subtraction. Section~\ref{sec:applications} presents the results from the SKA simulation and the Tianlai observational data. Finally, we provide a discussion and summary of our findings in Section~\ref{sec:dis} and \ref{sec:sum}.

\section{Methodology} \label{sec:method}

\subsection{Notations and Preliminaries}
To establish a rigorous mathematical framework for the tensor-based foreground subtraction analysis, we define the following notations and multilinear algebraic operations. Throughout this paper, scalars are denoted by lowercase letters (e.g., $a$), vectors by bold-face lowercase letters (e.g., $\bm{a}$), matrices by bold-face capitals (e.g., $\bm{A}$), and tensors by calligraphic letters (e.g., $\mathcal{A}$). For a third-order tensor $\mathcal{A} \in \mathbb{C}^{I_1 \times I_2 \times I_3}$, its $(i,j,k)$-th element is denoted by $a_{ijk}$. Key notations used in this work are summarized in Table~\ref{tab:notations}.

\begin{table}[ht]
\centering
\caption{Summary of Key Notations and Operations}
\label{tab:notations}
\begin{tabular}{ll}
\hline \hline
Notation & Definition/Description \\
\hline
$a, \bm{a}, \bm{A}, \mathcal{A}$ & Scalar, vector, matrix, and tensor \\
$a_{ijk}$ & $(i,j,k)$-th element of tensor $\mathcal{A}$ \\
$\mathcal{A}(:,:,k)$ & $k$-th frontal slice of a third-order tensor \\
$\bm{A}_{(n)}$ & Mode-$n$ unfolding (matricization) of $\mathcal{A}$ \\
$\mathbf{rank}_{n}(\mathcal{A})$ & Mode-$n$ rank of tensor $\mathcal{A}$ \\
$\mathcal{A} \times_n \bm{B}$ & Mode-$n$ product of tensor $\mathcal{A}$ and matrix $\bm{B}$ \\
$\mathcal{A} *_3 \mathcal{B}$ & Three-mode tensor-tensor product \\
$\langle \mathcal{A}, \mathcal{B} \rangle$ & Inner product of tensors $\mathcal{A}$ and $\mathcal{B}$ \\
$\|\mathcal{A}\|_F$ & Frobenius norm of tensor $\mathcal{A}$ \\
$\bm{a} \circ \bm{b} \circ \bm{c}$ & Vector outer product (rank-1 tensor) \\
\hline
\end{tabular}
\end{table}

A \textit{mode-$n$ fiber} is defined as a vector obtained by fixing all indices except for the $n$-th one, serving as the higher-order analogue of matrix rows and columns. A $k$-th \textit{frontal slice} of a third-order tensor $\mathcal{A}$ is denoted by $\mathcal{A}(:,:,k)$, representing a 2D matrix. The \textit{mode-$n$ unfolding} (also referred to as matricization) of a tensor $\mathcal{A}$, denoted by $\bm{A}_{(n)}$, is the process of reordering the mode-$n$ fibers into the columns of a matrix.

The \textit{$n$-rank} of a tensor $\mathcal{A}$, denoted by $\mathbf{rank}_{n}(\mathcal{A})$, is the dimension of the vector space spanned by its mode-$n$ fibers.  For example, $\mathbf{rank}_{3}(\mathcal{A}) = \mathbf{rank}(\bm{A}_{(3)})$.

The \textit{mode-$n$ product} of a tensor $\mathcal{A} \in \mathbb{C}^{I_1 \times I_2 \times \cdots \times I_N}$ with a matrix $\bm{B} \in \mathbb{C}^{J \times I_n}$, denoted by $\mathcal{A} \times_n \bm{B}$, is a tensor of size $I_1 \times \cdots \times I_{n-1} \times J \times I_{n+1} \times \cdots \times I_N$. Each mode-$n$ fiber of $\mathcal{A}$ is linearly transformed by the matrix $\bm{B}$. Element-wise, the operation is expressed as:
\begin{equation}
(\mathcal{A} \times_n \bm{B})_{i_1 \dots i_{n-1} j i_{n+1} \dots i_N} = \sum_{i_n=1}^{I_n} a_{i_1 \dots i_{n-1} i_n i_{n+1} \dots i_N} b_{j i_n}.
\end{equation}

The \textit{inner product} of two tensors $\mathcal{A}, \mathcal{B}$ of the same dimensions is defined as the sum of the products of their corresponding entries:
\begin{equation}
    \langle \mathcal{A}, \mathcal{B} \rangle = \sum_{i_1, i_2, \dots, i_N} \overline{a}_{i_1 i_2 \dots i_N} b_{i_1 i_2 \dots i_N},
\end{equation}
where the overline denotes complex conjugation. The \textit{Frobenius norm} of a tensor $\mathcal{A}$ is then given by $\|\mathcal{A}\|_F = \sqrt{\langle \mathcal{A}, \mathcal{A} \rangle}$.

A third-order \textit{rank-1 tensor} $\mathcal{A} \in \mathbb{C}^{I_1 \times I_2 \times I_3}$ is representable as the outer product of three vectors:
\begin{equation*}
   \mathcal{A}  =  \sigma (\bm{a} \circ \bm{b} \circ \bm{c}) \quad \iff \quad a_{ijk} = \sigma a_{i} b_{j} c_{k},
\end{equation*}
where $\sigma \in \mathbb{R}$ acts as a normalization scaling factor, and $\bm{a}$, $\bm{b}$, and $\bm{c}$ are unit vectors of lengths $I_1, I_2$, and $I_3$, respectively. In the language of mode-$n$ multiplication, this decomposition is concisely written as $\sigma \times_{1} \bm{a} \times_{2} \bm{b} \times_{3} \bm{c}$, where $\sigma$ is treated as a $1 \times 1 \times 1$ core tensor.

Finally, we introduce the \textit{tensor-tensor product} following \citep{Zeng2020}. For tensors $\mathcal{A} \in \mathbb{C}^{I_1 \times I_2 \times I_3}$ and $\mathcal{B} \in \mathbb{C}^{I_2 \times I_4 \times I_3}$, their \textit{three-mode} product $\mathcal{C} = \mathcal{A} *_3 \mathcal{B}$ is a tensor of size $I_1 \times I_4 \times I_3$ defined by slice-wise matrix multiplication:
\begin{equation}
\mathcal{C}(:,:,k) = \mathcal{A}(:,:,k) \mathcal{B}(:,:,k), \quad k=1,\dots,I_3.
\end{equation}
This operation effectively applies a sequence of linear transformations across the third dimension (the oriented axis), which is central to the O-SVD framework.

\subsection{Oriented Singular Value Decomposition (O-SVD)} \label{sec:osvd}
The Oriented Singular Value Decomposition (O-SVD) provides a hierarchical multilinear framework for decomposing third-order tensors by prioritizing a specific dimension—the oriented axis. While our exposition and subsequent applications primarily focus on 21\,cm intensity mapping (IM) datasets, it is important to emphasize that O-SVD is a mathematically general framework applicable to any third-order tensor where one dimension possesses distinct physical or statistical properties that warrant its selection as the oriented axis. For clarity of exposition in the context of 21\,cm IM, we demonstrate the method where this oriented axis corresponds to the spectral (frequency) dimension (Mode-3). This dimension is physically distinct from the spatial dimensions (Mode-1 and Mode-2) due to the high spectral coherence of astrophysical foreground emission, which typically exhibits smooth, power-law-like variations along the frequency axis, whereas the 21\,cm signal and thermal noise are characterized by rapid spectral fluctuations. The O-SVD framework exploits this physical asymmetry by prioritizing the spectral dimension as the oriented axis, while simultaneously capturing the spatial correlations through a nested SVD procedure. For a third-order tensor $\mathcal{A} \in \mathbb{C}^{I_{1} \times I_{2} \times I_{3}}$ with \textit{3-rank} $R_{3} = \mathbf{rank}_{3}(\mathcal{A})$, the O-SVD is defined as:
\begin{equation} \label{eq:osvd}
  \mathcal{A} = (\mathcal{U} *_{3} \mathcal{S} *_{3} \mathcal{V}) \times_{3} \bm{U}^{(3)},
\end{equation}
where $\bm{U}^{(3)} \in \mathbb{C}^{I_{3} \times I_{3}}$ is a unitary matrix obtained from the SVD of the mode-3 unfolding $\bm{A}_{(3)}$. The tensors $\mathcal{U} \in \mathbb{C}^{I_{1} \times I_{1} \times I_{3}}$, $\mathcal{S} \in \mathbb{C}^{I_{1} \times I_{2} \times I_{3}}$, and $\mathcal{V} \in \mathbb{C}^{I_{2} \times I_{2} \times I_{3}}$ satisfy the following properties:
\begin{enumerate}
    \item Each frontal slice $\mathcal{U}(:, :, k)$ and $\mathcal{V}(:, :, k)$ is a unitary matrix, and $\mathcal{S}(:, :, k)$ is a non-negative diagonal matrix for $k = 1, 2, \dots, R_{3}$;
    \item All frontal slices are null matrices for $k = R_{3} + 1, \dots, I_{3}$.
\end{enumerate}

The diagonal entries $s_{jjk}$ of the core tensor $\mathcal{S}$ represent the O-SVD singular values. These values obey a hierarchical ordering property:
\begin{equation*}
  \| \mathcal{S}(:, :, 1) \|_{F} \ge \| \mathcal{S}(:, :, 2) \|_{F} \ge \dots \ge \| \mathcal{S}(:, :, I_{3}) \|_{F} \ge 0,
\end{equation*}
where the Frobenius norm of each slice corresponds to the $k$-th singular value of the mode-3 unfolding, i.e., $\| \mathcal{S}(:, :, k) \|_{F} = \sigma_{k}$. Within each slice $k$, the singular values are further ordered as:
\begin{equation*}
  \sigma_{k} \ge s_{11k} \ge s_{22k} \ge \dots \ge s_{r_{2}r_{2}k} \ge 0,
\end{equation*}
for $k = 1, 2, \dots, r_{1}$, where $r_{1} = \min\{I_{3}, I_{1}I_{2}\}$ and $r_{2} = \min\{I_{1}, I_{2}\}$.

The computational procedure involves two primary stages. First, a standard matrix SVD is performed on the $I_{3} \times I_{1}I_{2}$ mode-3 unfolding matrix $\bm{A}_{(3)}$:
\begin{equation*}
  \bm{A}_{(3)} = \bm{U}^{(3)} \bm{\Sigma}^{(3)} \bm{V}^{(3)H},
\end{equation*}
where $\bm{U}^{(3)}$ is the unitary matrix of spectral basis vectors, and $\bm{V}^{(3)}$ contains the corresponding flattened spatial modes. The singular values $\sigma_{k}$ of $\bm{A}_{(3)}$ are the diagonal elements of $\bm{\Sigma}^{(3)}$. In the second stage, each spatial mode $\bm{v}_{k}$ (the $k$-th column of $\bm{V}^{(3)}$) is reshaped into an $I_1 \times I_2$ matrix $\tilde{\bm{V}}_k$ and decomposed via SVD:
\begin{equation} \label{eq:Vksvd}
  \tilde{\bm{V}}_{k} = \bm{U}_{k} \bm{\Sigma}_{k} \bm{V}_{k}^{H},
\end{equation}
where $\bm{\Sigma}_{k}$ contains the spatial singular values $\sigma_{kj}$. The relationship between the O-SVD singular values and the matrix singular values is given by $s_{jjk} = \sigma_k \sigma_{kj}$, which implies:
\begin{equation} \label{eq:ss}
  \sigma_{k}^{2} = \sum_{j=1}^{r_{2}} s_{jjk}^{2}.
\end{equation}
This derivation follows from the property that the Frobenius norm is invariant under unitary transformations: $\sum_{j=1}^{r_{2}} \sigma_{kj}^{2} =  \|\tilde{\bm{V}}_k\|_F^2 = \|\bm{V}^{(3)}(:, k)\|_2^2 = 1$.

The two-stage O-SVD process is illustrated in Figure~\ref{fig:osvd_detailed} for an example $3 \times 3 \times 3$ tensor.

\begin{figure*}[htbp]
\centering
\begin{tikzpicture}[
    >=stealth,
    scale=0.85,
    every node/.style={scale=0.85},
    matrix/.style={
        rectangle,
        draw=black,
        thick,
        minimum width=2.5cm,
        minimum height=1.2cm,
        align=center
    },
    box/.style={
        rectangle,
        draw=black,
        thick,
        fill=#1,
        minimum width=1.5cm,
        minimum height=1.0cm,
        align=center
    },
    arrow/.style={
        thick,
        ->,
        shorten >=2pt,
        shorten <=2pt
    },
    stage/.style={
        font=\bfseries\large,
        align=center
    }
]

% ==========
% Stage 1 Title
% ==========
\node at (-3, 6) [stage] {Stage 1};

% ==========
% Original tensor notation
% ==========
\node at (-0.3, 5) {$\mathcal{A} \in \mathbb{C}^{I_1 \times I_2 \times I_3}$};

% ==========
% Tensor visualization with slices
% ==========
\node at (-0.4, 3.5) {
$
\begin{bmatrix}
\begin{bmatrix}
a_{111} & a_{121} & a_{131} \\
a_{211} & a_{221} & a_{231} \\
a_{311} & a_{321} & a_{331}
\end{bmatrix}
\begin{bmatrix}
a_{112} & a_{122} & a_{132} \\
a_{212} & a_{222} & a_{232} \\
a_{312} & a_{322} & a_{332}
\end{bmatrix}
\begin{bmatrix}
a_{113} & a_{123} & a_{133} \\
a_{213} & a_{223} & a_{233} \\
a_{313} & a_{323} & a_{333}
\end{bmatrix}
\end{bmatrix}
$
};

% ==========
% Arrow to unfolded matrix
% ==========
\draw[arrow] (4.2, 3.5) -- (5.8, 3.5);

% ==========
% Unfolded matrix
% ==========
\node at (10, 5) {$\mathbf{A}_{(3)} \in \mathbb{C}^{I_3 \times I_1 I_2}$};

\node at (10, 3.5) {
$
\begin{bmatrix}
a_{111} & a_{121} & a_{131} & a_{211} & a_{221} & a_{231} & a_{311} & a_{321} & a_{331} \\
a_{112} & a_{122} & a_{132} & a_{212} & a_{222} & a_{232} & a_{312} & a_{322} & a_{332} \\
a_{113} & a_{123} & a_{133} & a_{213} & a_{223} & a_{233} & a_{313} & a_{323} & a_{333}
\end{bmatrix}
$
};

% ==========
% Arrow down to SVD
% ==========
\draw[arrow] (10, 2.5) -- (10, 1.2);

% ==========
% SVD equation
% ==========
\node at (10, 0.8) {$\mathbf{A}_{(3)} = \mathbf{U}^{(3)} \mathbf{\Sigma}^{(3)} \mathbf{V}^{(3)H}$};

% ==========
% SVD matrices
% ==========
\node at (2.8, 0) {$\mathbf{U}^{(3)}$};

\node at (2.8, -1.5) {
$
\begin{bmatrix}
u^{(3)}_{11} & u^{(3)}_{12} & u^{(3)}_{13} \\
u^{(3)}_{21} & u^{(3)}_{22} & u^{(3)}_{23} \\
u^{(3)}_{31} & u^{(3)}_{32} & u^{(3)}_{33}
\end{bmatrix}
$
};

\node at (4.3, -1.5) {$\times$};

\node at (5.5, 0) {$\mathbf{\Sigma}^{(3)}$};

\node at (5.5, -1.5) {
$
\begin{bmatrix}
\sigma_1 & 0 & 0 \\
0 & \sigma_2 & 0 \\
0 & 0 & \sigma_3
\end{bmatrix}
$
};

\node at (6.7, -1.5) {$\times$};

\node at (11, 0) {$\mathbf{V}^{(3)H}$};

\node at (11, -1.5) {
$
\begin{bmatrix}
v^{(3)*}_{11} & v^{(3)*}_{21} & v^{(3)*}_{31} & v^{(3)*}_{41} & v^{(3)*}_{51} & v^{(3)*}_{61} & v^{(3)*}_{71} & v^{(3)*}_{81} & v^{(3)*}_{91} \\
v^{(3)*}_{12} & v^{(3)*}_{22} & v^{(3)*}_{32} & v^{(3)*}_{42} & v^{(3)*}_{52} & v^{(3)*}_{62} & v^{(3)*}_{72} & v^{(3)*}_{82} & v^{(3)*}_{92} \\
v^{(3)*}_{13} & v^{(3)*}_{23} & v^{(3)*}_{33} & v^{(3)*}_{43} & v^{(3)*}_{53} & v^{(3)*}_{63} & v^{(3)*}_{73} & v^{(3)*}_{83} & v^{(3)*}_{93}
\end{bmatrix}
$
};

% ==========
% Stage 2 Title
% ==========
\node at (-3, -4) [stage] {Stage 2};

\node at (-1.8, -5) {$\mathbf{V}^{(3)} \in \mathbb{C}^{I_1 I_2 \times I_3}$};

\node at (-1.8, -7) {
$
\begin{bmatrix}
v^{(3)}_{11} & v^{(3)}_{12} & \dots \\
v^{(3)}_{21} & v^{(3)}_{22} & \dots \\
v^{(3)}_{31} & v^{(3)}_{32} & \dots \\
\vdots & \vdots & \ddots \\
v^{(3)}_{91} & v^{(3)}_{92} & \dots \\
\end{bmatrix}
$
};

%

% ==========
% Arrow to reshape
% ==========
\draw[arrow] (-0.1, -7) -- (1.5, -7);

% ==========
% Reshaped matrices
% ==========
\node at (5.0, -5) {
$
\begin{bmatrix}
\tilde{\mathbf{V}}_1,
\tilde{\mathbf{V}}_2,
\dots
\end{bmatrix}
$
};

\node at (5.0, -7) {
$
\begin{bmatrix}
\begin{bmatrix}
v^{(3)}_{11} & v^{(3)}_{21} & v^{(3)}_{31} \\
v^{(3)}_{41} & v^{(3)}_{51} & v^{(3)}_{61} \\
v^{(3)}_{71} & v^{(3)}_{81} & v^{(3)}_{91}
\end{bmatrix},
\begin{bmatrix}
v^{(3)}_{12} & v^{(3)}_{22} & v^{(3)}_{32} \\
v^{(3)}_{42} & v^{(3)}_{52} & v^{(3)}_{62} \\
v^{(3)}_{72} & v^{(3)}_{82} & v^{(3)}_{92}
\end{bmatrix},
\dots
\end{bmatrix}
$
};

% ==========
% Arrow to final decomposition
% ==========
\draw[arrow] (8.5, -7) -- (10.1, -7);

% ==========
% Final decomposition for each column
% ==========
\node at (12.5, -7) {
$
\begin{bmatrix}
\mathbf{U}_1 \mathbf{\Sigma}_1 \mathbf{V}_1^H,
\mathbf{U}_2 \mathbf{\Sigma}_2 \mathbf{V}_2^H,
\dots
\end{bmatrix}
$
};

% ==========
% Arrow down to SVD
% ==========
\draw[arrow] (12.5, -8.5) -- (12.5, -9.8);

% ==========
% SVD equation
% ==========
\node at (12.5, -10.3) {$\tilde{\mathbf{V}}_k = \mathbf{U}_k \mathbf{\Sigma}_k \mathbf{V}_k^H$};

% ==========
% SVD matrices
% ==========
\node at (6.3, -11.3) {$\mathbf{U}_k$};

% ==========
% Detailed decomposition example
% ==========
\node at (6.3, -12.6) {
$
\begin{bmatrix}
(\mathbf{u}_k)_{11} & (\mathbf{u}_k)_{12} & (\mathbf{u}_k)_{13} \\
(\mathbf{u}_k)_{21} & (\mathbf{u}_k)_{22} & (\mathbf{u}_k)_{23} \\
(\mathbf{u}_k)_{31} & (\mathbf{u}_k)_{32} & (\mathbf{u}_k)_{33}
\end{bmatrix}
$
};

\node at (8.4, -12.6) {$\times$};

\node at (9.8, -11.3) {$\mathbf{\Sigma}_k$};

\node at (9.8, -12.6) {
$
\begin{bmatrix}
\sigma_{k1} & 0 & 0 \\
0 & \sigma_{k2} & 0 \\
0 & 0 & \sigma_{k3}
\end{bmatrix}
$
};

\node at (11.2, -12.6) {$\times$};

\node at (13.2, -11.3) {$\mathbf{V}_k^H$};

\node at (13.2, -12.6) {
$
\begin{bmatrix}
(\mathbf{v}_k)_{11}^* & (\mathbf{v}_k)_{21}^* & (\mathbf{v}_k)_{31}^* \\
(\mathbf{v}_k)_{12}^* & (\mathbf{v}_k)_{22}^* & (\mathbf{v}_k)_{32}^* \\
(\mathbf{v}_k)_{13}^* & (\mathbf{v}_k)_{23}^* & (\mathbf{v}_k)_{33}^*
\end{bmatrix}
$
};

\end{tikzpicture}
\caption{Oriented Singular Value Decomposition (O-SVD) two-stage process. Stage 1 unfolds the tensor and performs SVD on the mode-3 matrix. Stage 2 takes each column of $\mathbf{V}^{(3)}$, reshapes it into an $I_1 \times I_2$ matrix, and performs SVD on each reshaped matrix.}
\label{fig:osvd_detailed}
\end{figure*}

This hierarchical structure allows $\mathcal{A}$ to be expressed as a superposition of rank-1 outer products:
\begin{equation} \label{eq:Ar1}
  \mathcal{A} = \sum_{k=1}^{r_{1}} \sum_{j=1}^{r_{2}} s_{jjk} \times_{1} \bm{u}_{kj} \times_{2} \bm{v}_{kj} \times_{3} \bm{u}_{k},
\end{equation}
where $\bm{u}_{k}$ represents the spectral basis vector, while $\bm{u}_{kj}$ and $\bm{v}_{kj}$ represent the spatial basis vectors. This decomposition provides a more refined representation than the matrix SVD by explicitly separating spatial and spectral degrees of freedom. The complete computational process is summarized in Algorithm~\ref{alg:osvd}.

It is worth highlighting the intrinsic connection between O-SVD and the standard Principal Component Analysis (PCA) widely adopted in 21\,cm cosmology. PCA is mathematically equivalent to performing an SVD on the mode-3 unfolded matrix $\bm{A}_{(3)}$. In this context, the first stage of the O-SVD procedure is identical to PCA, capturing the dominant spectral variations. However, while PCA treats the associated spatial modes as flattened 1D vectors, O-SVD further decomposes these modes into their constituent spatial basis vectors. This hierarchical structure positions O-SVD as a direct ``plug-and-play'' generalization of PCA: any foreground mitigation pipeline currently employing PCA can be upgraded to O-SVD by replacing the matrix SVD step. This transition allows the pipeline to leverage additional spatial-spectral filtering degrees of freedom without requiring fundamental changes to the data processing architecture.

For foreground subtraction, we leverage the concentration of foreground power in the largest singular values. By truncating the O-SVD expansion at $r$ terms, we obtain a low-rank approximation $\mathcal{A}_r$ that minimizes the Frobenius norm of the residual:
\begin{equation} \label{eq:Ar}
  \|\mathcal{A} - \mathcal{A}_{r}\|_{F}^{2} = \sum_{i=r + 1}^{r_{1} r_{2}} \tilde{s}_{i}^{2},
\end{equation}
where $\tilde{s}_i$ are the reordered O-SVD singular values in decreasing order. This approach allows for a more flexible and precise isolation of foreground modes compared to traditional PCA-based methods, as the O-SVD offers additional degrees of freedom to characterize the complex spatio-spectral morphologies of astrophysical foregrounds.

\section{Applications} \label{sec:applications}

\subsection{Simulated SKA Data} \label{subsec:ska}
We evaluate the performance of the O-SVD method using the SKA Science Data Challenge 3a (SDC3a, \url{https://sdc3.skao.int/challenges/foregrounds}) dataset, which was established by the Square Kilometre Array Observatory (SKAO) as a standardized benchmark for Epoch of Reionization (EoR) foreground mitigation. A preliminary description of the O-SVD approach and its performance during the challenge (as submitted by the Shuimu-Tianlai team) is provided in Section 3.14 of \citet{2025MNRAS.543.1092B}. The method demonstrated significant efficacy in foreground subtraction, as characterized by its competitive rankings and detailed statistical validation in that work. In this study, we utilize the realistic simulation environment of SDC3a as a controlled testbed to further investigate the theoretical and practical advantages of the O-SVD framework, moving beyond the initial challenge results to provide a more comprehensive analysis of signal recovery. The source code for the analysis pipeline is publicly available on GitHub: \url{https://github.com/zuoshifan/sdc3a_osvd_pipeline}

\subsubsection{The SDC3a Dataset}
Our analysis utilizes the SKA Science Data Challenge 3a (SDC3a) dataset, which represents a high-fidelity and comprehensive simulation of SKA-Low observations targeting the Epoch of Reionization (EoR) \citep{2025arXiv250609533B}. Designed to simulate complex observational conditions, the dataset provides a rigorous testbed for evaluating the efficacy of foreground mitigation algorithms in the presence of realistic instrumental systematics and dominant astrophysical emissions.

The simulation spans a frequency range of 106--196\,MHz, corresponding to the redshift interval $z \approx 6.2$--$12.4$. The data comprises 900 spectral channels with a frequency resolution of 100\,kHz. The simulated components are categorized as follows:
\begin{itemize}
    \item \textbf{Astrophysical Foregrounds}: Foreground emissions exceed the cosmological signal by several orders of magnitude.
    \begin{itemize}
        \item \textit{Galactic Emission}: Diffuse Galactic synchrotron radiation modeled after a modified version of the Global Sky Model (GSM2016; \citealt{2017MNRAS.464.3486Z}), with quadratic interpolation in log(frequency) and additional spatial frequency content beyond the native resolution from synthetic observations \citep{2025arXiv250609533B}.
        \item \textit{Extragalactic Sources}: This component includes high-flux ``A-Team'' sources ($>5$\,Jy at 150\,MHz) and a dense population of fainter sources derived from the GLEAM and LoBES catalogs ($>100$\,mJy at 150\,MHz) \citep{2021PASA...38...57L}. Additionally, mock continuum simulations (T-RECS) \citep{2019MNRAS.482....2B,2023MNRAS.524..993B} were used to model the sub-mJy population down to $1\,\mu\text{Jy}$.
    \end{itemize}
    \item \textbf{Cosmological Signal}: The redshifted 21\,cm signal was generated using \texttt{21cmFAST} \citep{2011MNRAS.411..955M,2020JOSS....5.2582M}, providing the 3D brightness temperature fluctuations that constitute the target signal for recovery.
    \item \textbf{Instrumental Systematics}: The simulation incorporates realistic SKA-Low telescope response effects:
    \begin{itemize}
        \item \textit{Residual Point Sources}: Partially successful ``de-mixing'' of far-sidelobe sources was emulated by including out-of-field sources attenuated by a factor of $10^{-3}$.
        \item \textit{Ionospheric Effects}: Modeled using \texttt{ARATMOSPY} \citep{2015OExpr..2333335S} with a correlation scale of $r_0 = 7$\,km, assuming moderately successful direction-dependent calibration (attenuation factor of $10^{-2}$).
        \item \textit{Calibration Errors}: Direction-independent gain errors were modeled as Gaussian noise in both phase (0.02$^\circ$) and amplitude (0.02\%) across time and frequency.
        \item \textit{Thermal Noise}: Instrumental noise was scaled to represent an average of two polarizations and a total integration time of 1000 hours, consistent with nominal SKA-Low sensitivity profiles provided by \texttt{OSKAR}.
    \end{itemize}
\end{itemize}

The objective of the challenge was the accurate recovery of the cylindrical 2D power spectrum $P(k_\perp, k_\parallel)$ of the 21\,cm signal. The SDC3a data products include gridded visibilities (Measurement Set and UVFITS formats, $\sim 7.5$\,TB) produced by \texttt{OSKAR} \citep{2009wska.confE..31D}, as well as synthesized image and point spread function (PSF) cubes generated using \texttt{WSCLEAN} \citep{2014MNRAS.444..606O}.

\subsubsection{Data Processing} \label{S:dpa}
For our analysis, we specifically utilized the ``uniform-weighted'' image cubes, which provide superior angular resolution and more symmetric beam patterns compared to natural weighting, thereby facilitating the isolation of point-like extragalactic foregrounds. Following the SDC3a submission requirements, the 90\,MHz total bandwidth is partitioned into six sub-bands of 15\,MHz each, with independent power spectra computed for each interval. To mitigate edge effects and suppress noise, the analysis is restricted to the central $4^{\circ} \times 4^{\circ}$ field of view. For the purpose of demonstrating the O-SVD framework's efficacy in foreground mitigation, we present detailed results for the first sub-band (106--121\,MHz), which represents the most challenging regime due to the high foreground-to-signal ratio. The analysis pipeline was implemented using a suite of specialized astrophysical software packages, as described below.

\textit{Unit and Coordinate Transformation:} The raw image cubes, initially calibrated in units of Jy/beam, were converted to brightness temperature ($T_b$) in Kelvin. This conversion was performed using the \texttt{radio\_beam}\footnote{\url{https://pypi.org/project/radio-beam/}} package, which facilitates the extraction of frequency-dependent synthesized beam parameters from FITS headers to ensure consistent flux scaling across the entire bandwidth.

\begin{figure*}
\centering
\begin{minipage}{0.38\linewidth}
\centering\includegraphics[width=\textwidth]{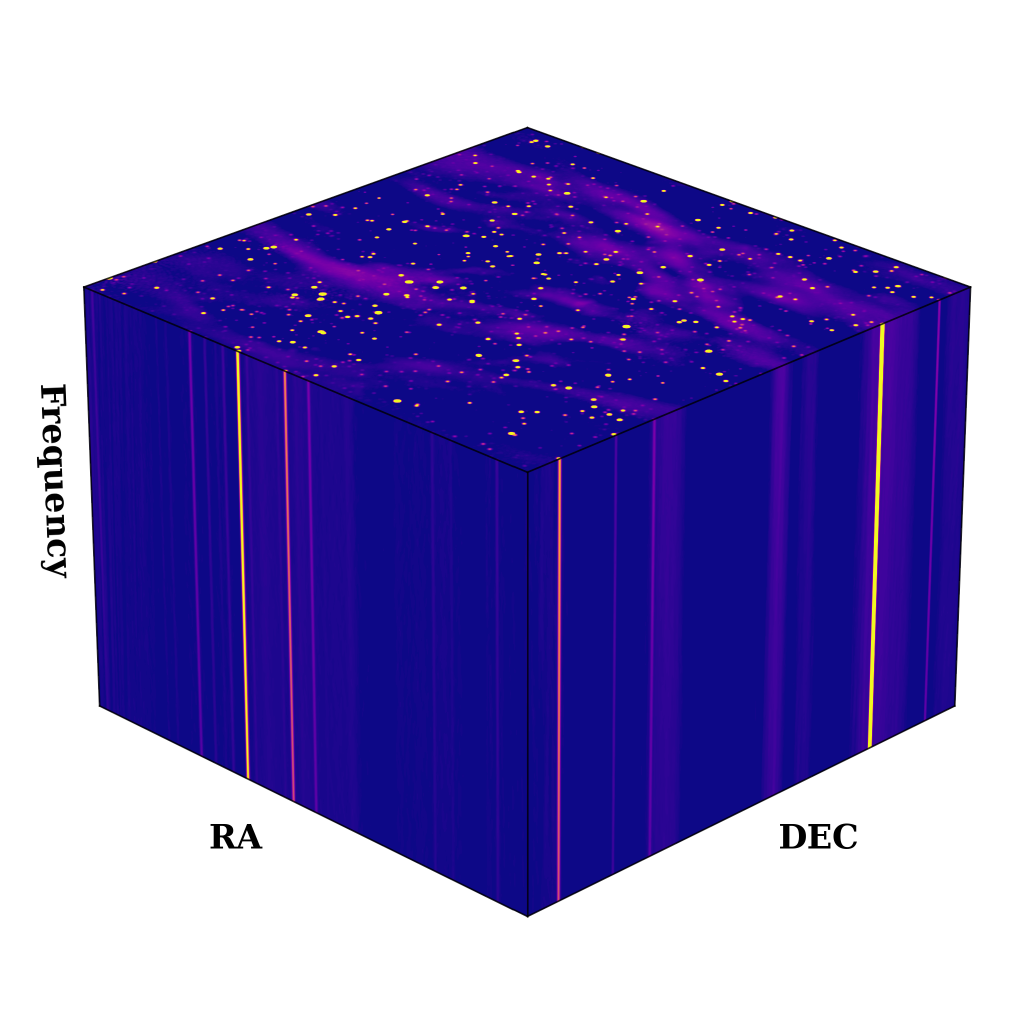}
\end{minipage}
\begin{minipage}{0.57\linewidth}
\centering\includegraphics[width=\textwidth]{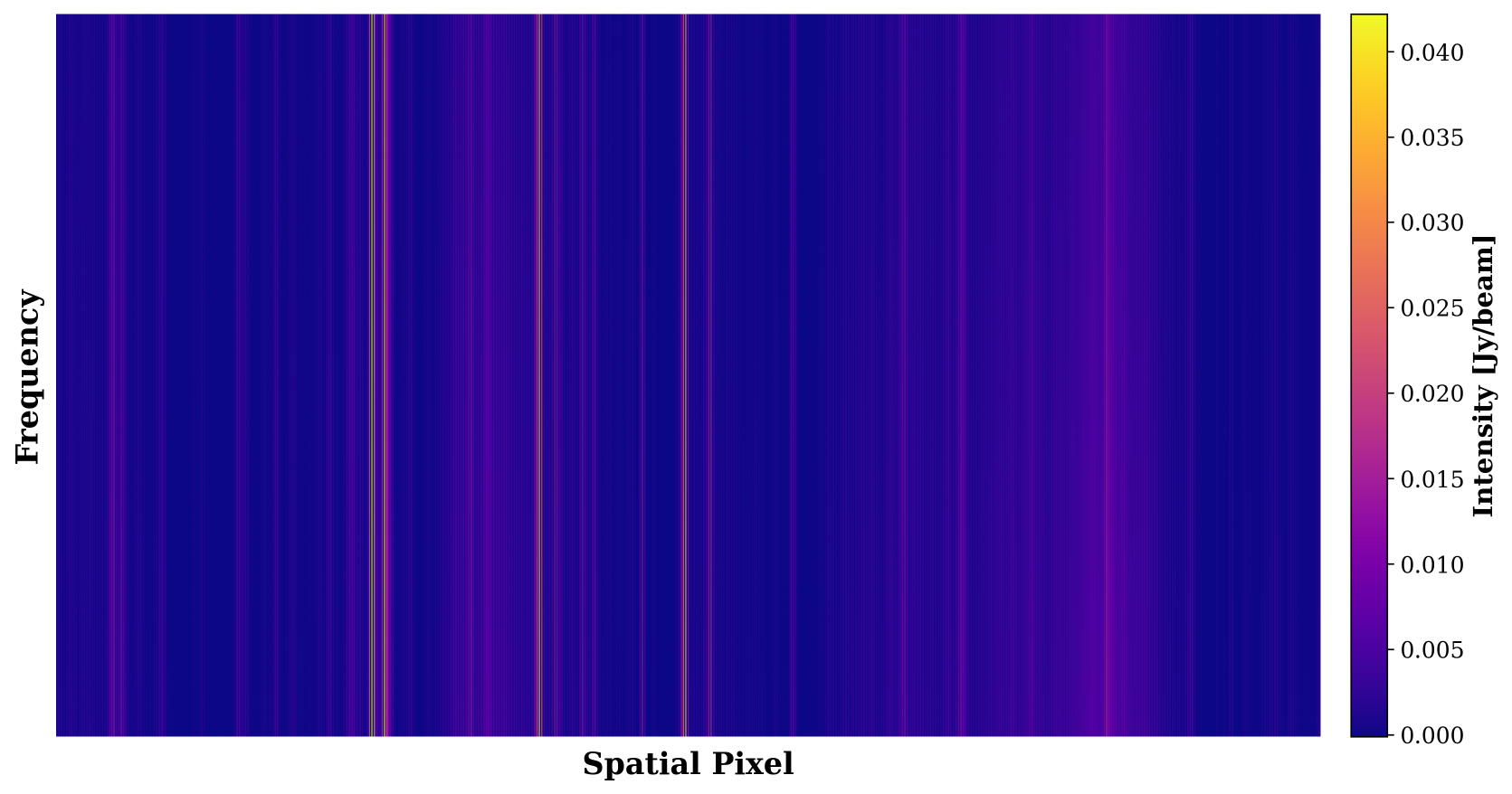}
\end{minipage}
\caption{The first sub-band (106--121\,MHz) of the SKA SDC3a simulation visualized as a 3D image cube (left) and as a flattened 2D array by combining the two spatial axes (right) with identical color scaling. The 3D representation preserves the intrinsic spatial-spectral topology, whereas flattening collapses the spatial information into a single dimension, potentially obscuring the multi-dimensional correlations utilized by O-SVD.}
\label{fig:imcube}
\end{figure*}

\textit{Tensor Formulation and Foreground Mitigation:} The processed image cube was formalized as a third-order tensor $\mathcal{T} \in \mathbb{R}^{N_x \times N_y \times N_{\nu}}$, with $N_x = N_y = 900$ representing spatial dimensions and $N_{\nu} = 150$ representing the spectral channels. The left panel of Figure~\ref{fig:imcube} visualizes this image cube. We applied the O-SVD method to decompose $\mathcal{T}$ into its hierarchical multilinear components. Given that foreground emissions typically exhibit high spectral smoothness and spatial coherence, they are preferentially concentrated in the dominant modes associated with the largest O-SVD singular values. A truncation threshold $N_{fg}$ was determined by identifying the transition point in the singular value spectrum where the steeply declining foreground components reach the noise-dominated floor (detailed below). The foreground-subtracted tensor $\mathcal{T}_{\text{res}}$, containing the estimated 21\,cm signal and instrumental noise, was then obtained by subtracting the first $N_{fg}$ modes.

\begin{figure*}
\centering
\begin{minipage}{0.45\linewidth}
\centering\includegraphics[width=\textwidth]{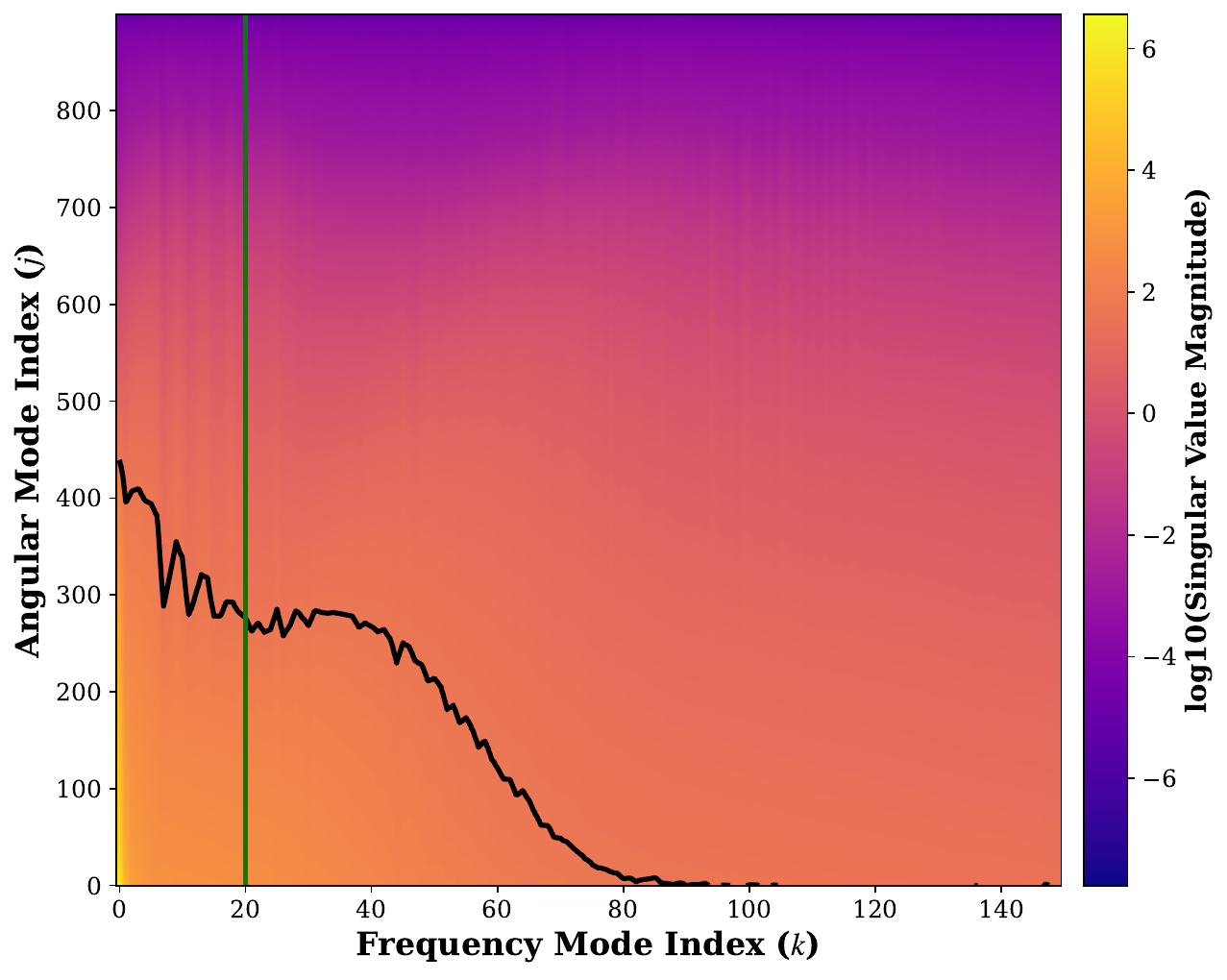}
\end{minipage}
\begin{minipage}{0.45\linewidth}
\centering\includegraphics[width=\textwidth]{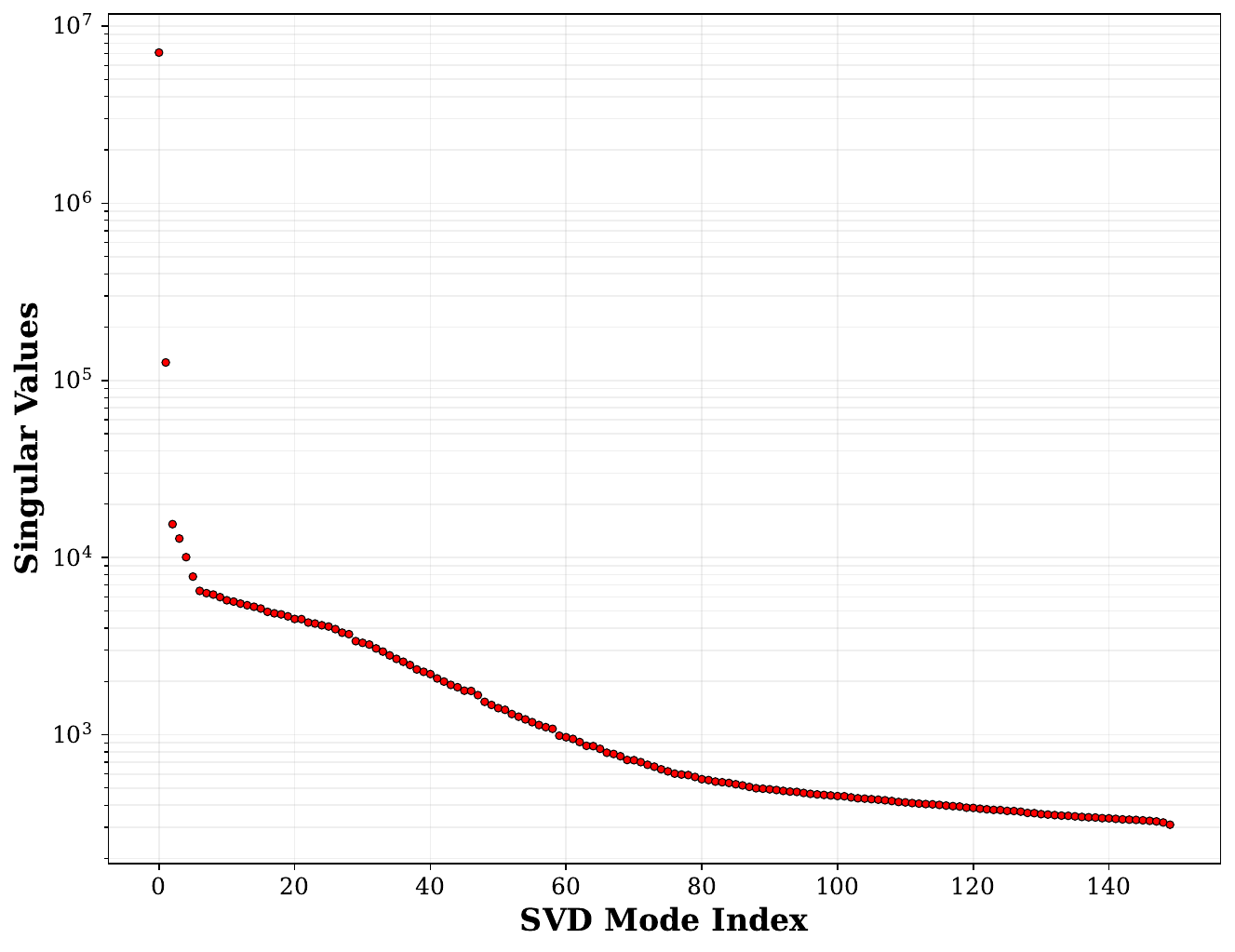}
\end{minipage}
\caption{Comparison of singular value spectra: (Left) The 2D O-SVD singular value array $s_{jjk}$, where the indices $j$ and $k$ correspond to the spatial and spectral modes, respectively. The black contour indicates the truncation threshold $s_{11, k=100}$ used for foreground subtraction. The green line indicates the truncation threshold for $N_{fg}^{\rm PCA} = 20$ used for traditional SVD. (Right) The 1D matrix SVD singular value spectrum $\sigma_k$ obtained from the flattened 2D array.}
\label{fig:sigvals}
\end{figure*}

\textit{Coordinate Transformation:} The angular coordinates $(\theta_x, \theta_y)$ and frequency $\nu$ were mapped to comoving physical distances $(L_x, L_y, L_z)$ in units of comoving Mpc. This transformation assumed a standard \textit{FlatLambdaCDM} cosmology (\texttt{astropy.cosmology}\footnote{\url{https://docs.astropy.org/en/stable/cosmology/index.html}}), with parameters $H_0 = 100$\,km\,s$^{-1}$\,Mpc$^{-1}$ and $\Omega_m = 0.30964$, consistent with the SDC3a requirements for comparative analysis against the ground truth.

\textit{Power Spectrum Estimation and Statistical Analysis:} The cylindrical 2D power spectrum $P(k_\perp, k_\parallel)$ was estimated from the residual tensor $\mathcal{T}_{\text{res}}$ using the \texttt{tools21cm}\footnote{\url{https://github.com/sambit-giri/tools21cm/}} library. This involved 3D Fourier transformation followed by cylindrical binning in $(k_\perp, k_\parallel)$ space. The efficacy of foreground suppression was quantified by comparing the residual power spectrum against the true EoR 21\,cm power spectrum released post-challenge.

\subsubsection{Analysis}
In Figure~\ref{fig:imcube}, the sub-band image cube is shown in the left panel, while the corresponding flattened 2D array obtained by collapsing the angular axes (i.e., reshaping the two angular axes into a single dimension) is shown in the right panel. As can be seen from this figure, the flattening process destroys the inherent spatial structural information.

Figure~\ref{fig:sigvals} illustrates the O-SVD singular values $s_{jjk}$ of the image cube (left) and the matrix SVD singular values $\sigma_{k}$ of the flattened 2D array (right). Following the relationship in Eq.~\ref{eq:ss}, each $\sigma_{k}$ in the matrix SVD corresponds to the quadratic sum of the $k$-th spectral mode's spatial singular values in the O-SVD.

\begin{figure}
\centering
\includegraphics[width=0.45\textwidth]{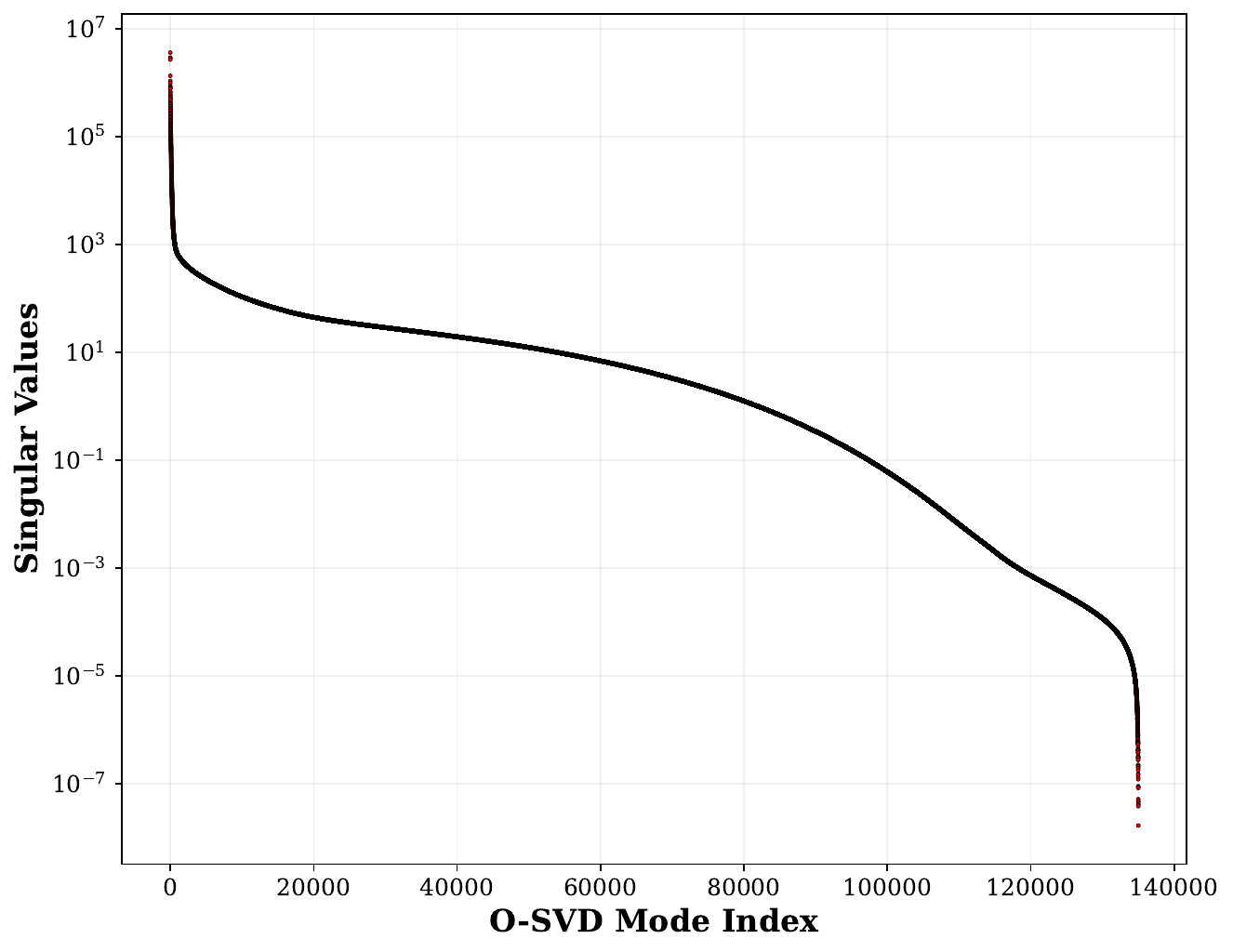}
\caption{The O-SVD singular values $s_{jjk}$ reordered in descending order.}
\label{fig:osvdsigv}
\end{figure}

To perform foreground subtraction, we truncate the O-SVD expansion by removing modes associated with the largest singular values. Figure~\ref{fig:osvdsigv} shows the O-SVD singular values reordered in descending sequence.

\begin{figure}
\centering
\begin{minipage}{0.9\linewidth}
\centering\includegraphics[width=\textwidth]{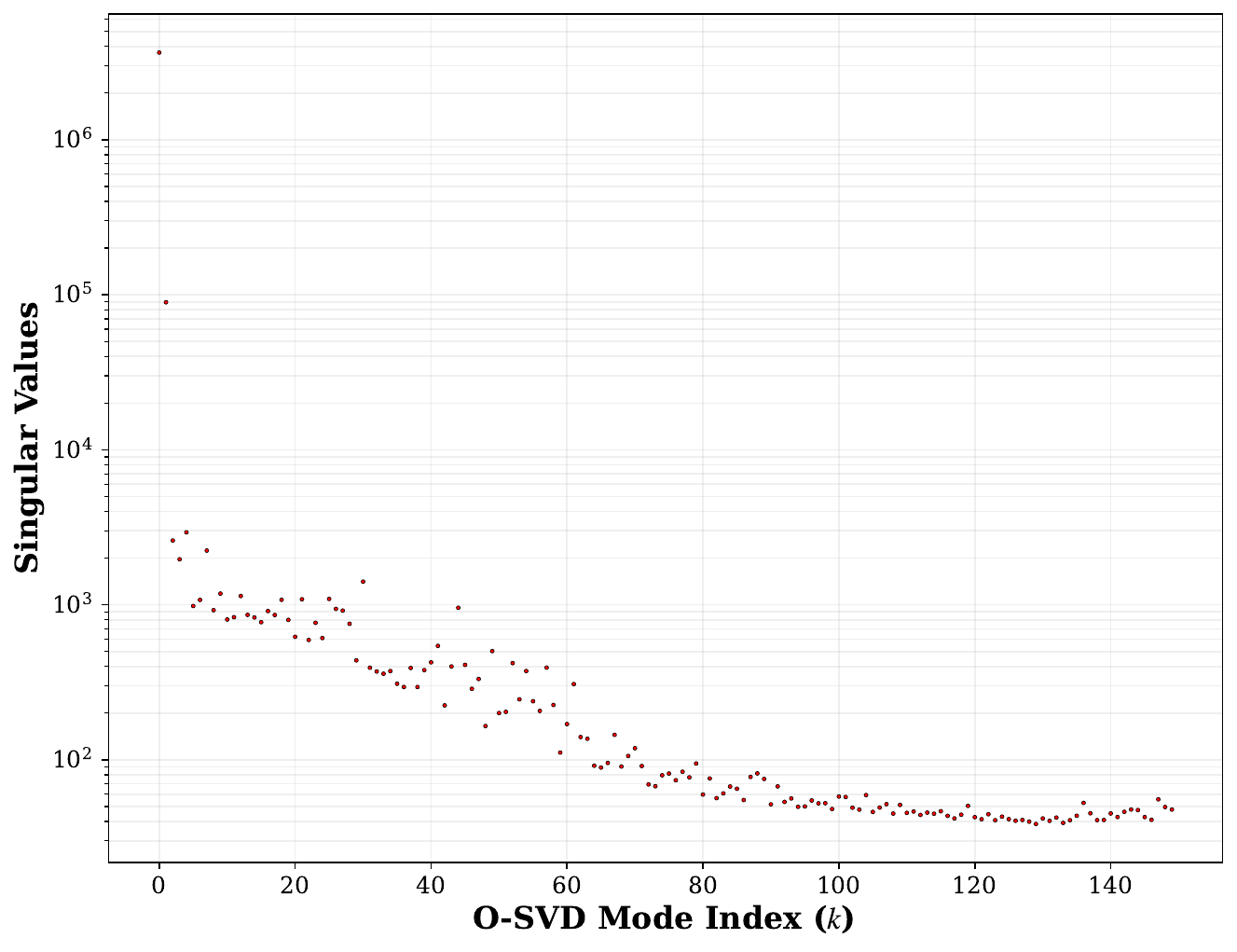}
\end{minipage}\\
\begin{minipage}{0.9\linewidth}
\centering\includegraphics[width=\textwidth]{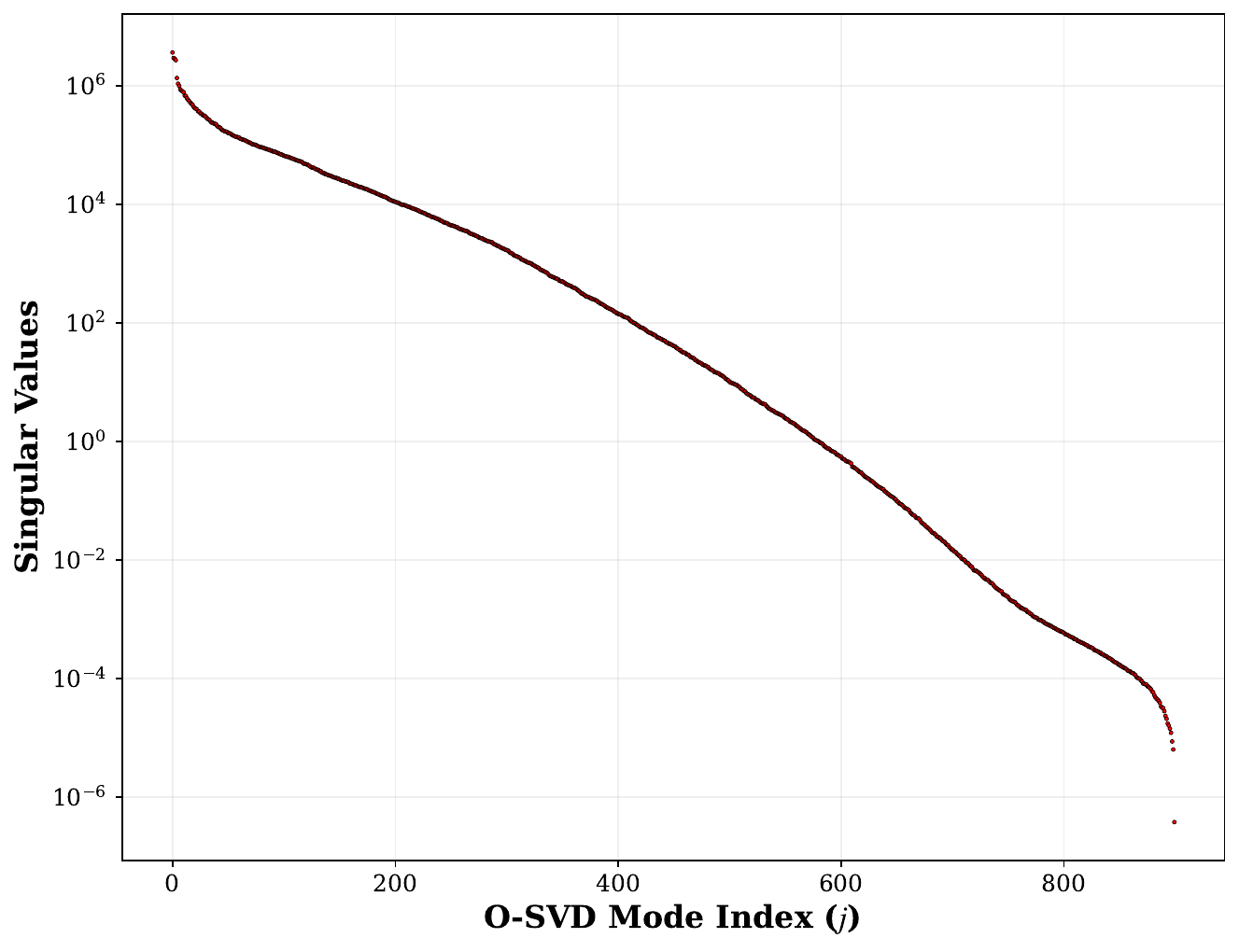}
\end{minipage}
\caption{The first row (spectral trend, $s_{11k}$, top) and first column (spatial trend, $s_{j11}$, bottom) of the O-SVD singular value array $s_{jjk}$. The spectral trend captures the variance across frequency slices, while the spatial trend reflects the hierarchical importance of spatial features within the primary spectral mode.}
\label{fig:sjk}
\end{figure}

\begin{figure}
\centering
\includegraphics[width=0.45\textwidth]{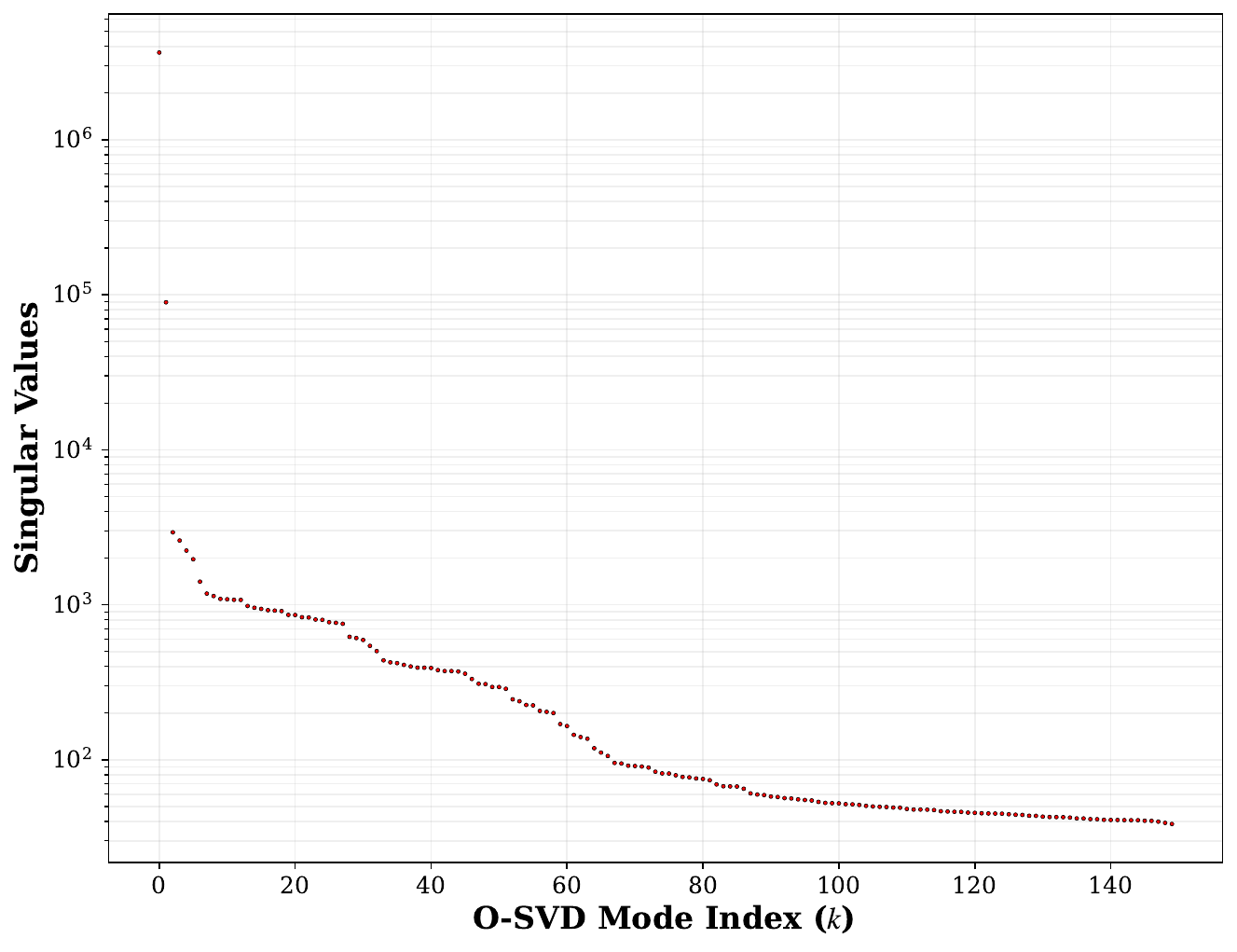}
\caption{The first row of the singular value array ($s_{11k}$) reordered in descending order. The plateau at large indices serves as an empirical diagnostic for identifying the noise floor and setting the truncation threshold.}
\label{fig:sk}
\end{figure}

Determining an optimal truncation threshold from the 1D reordered spectrum in Figure~\ref{fig:osvdsigv} is challenging, as there is no distinct gap between the foreground-dominated and noise-dominated regimes. However, the intrinsic 2D structure of the O-SVD singular values provides additional diagnostic information. Figure~\ref{fig:sjk} plots the primary spectral singular values $s_{11k}$ (top) and the primary spatial singular values $s_{jj1}$ (bottom). While $s_{jj1}$ is monotonically non-increasing by definition, $s_{11k}$ exhibits a general downward trend with local fluctuations. Notably, for $k \gtrsim 100$, the $s_{11k}$ values stabilize into an approximately flat floor, as shown more clearly in the reordered plot in Figure~\ref{fig:sk}. Consequently, we select $s_{11, k=100}$ as the truncation threshold, removing all modes satisfying $s_{jjk} \ge s_{11, k=100}$. These modes are localized in the lower-left region of the 2D singular value array (indicated by the black contour in Figure~\ref{fig:sigvals}, left).

This criterion yields $N_{fg} = 17,652$ removed modes, which is slightly higher than the $N'_{fg} = 15,000$ used in the initial SDC3a submission \citep{2025MNRAS.543.1092B}. While the previous threshold was determined through empirical visual comparison with foreground-free simulations, the O-SVD-based criterion presented here is more robust and applicable to real data where no such reference exists. The slightly higher mode count is expected for this sub-band (106--121\,MHz), which contains the strongest foreground emission.

\begin{figure*}
\centering
\begin{minipage}{0.4\linewidth}
\centering\includegraphics[width=\textwidth]{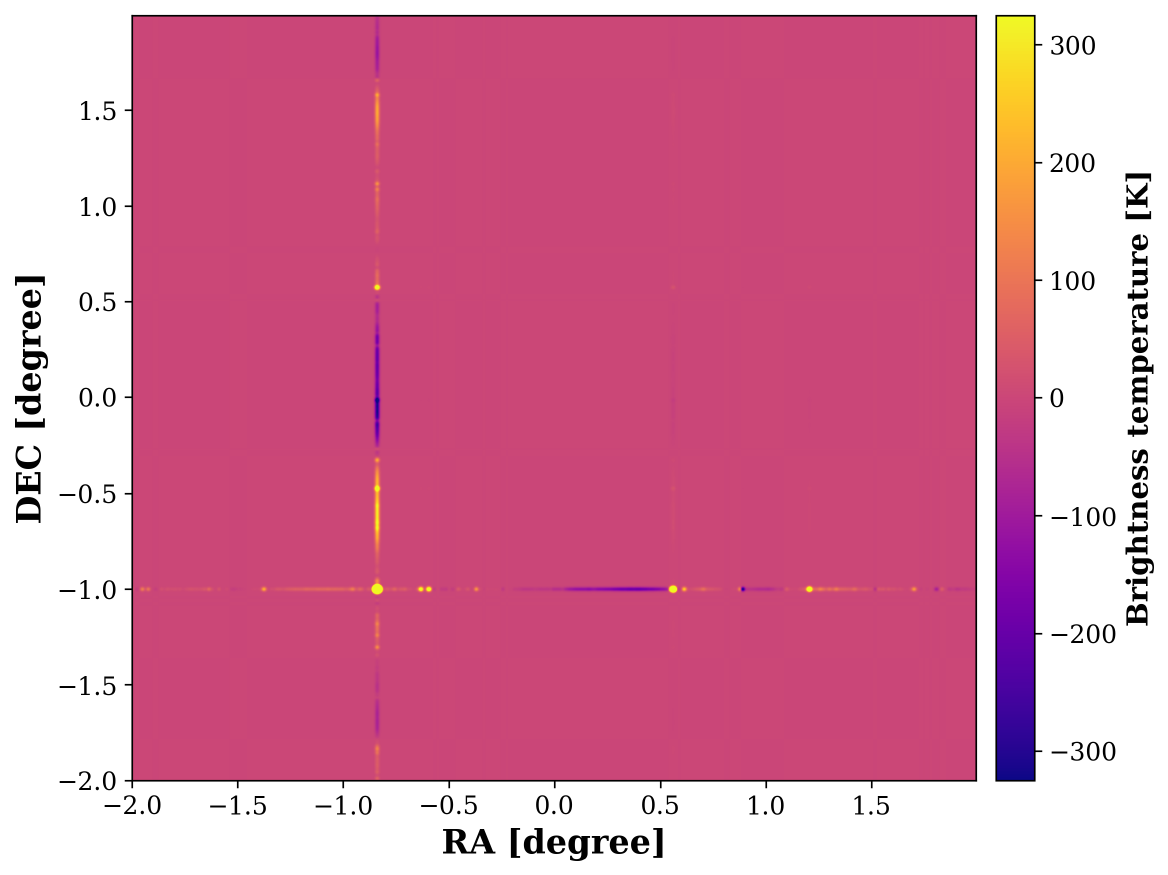}
\end{minipage}
\begin{minipage}{0.4\linewidth}
\centering\includegraphics[width=\textwidth]{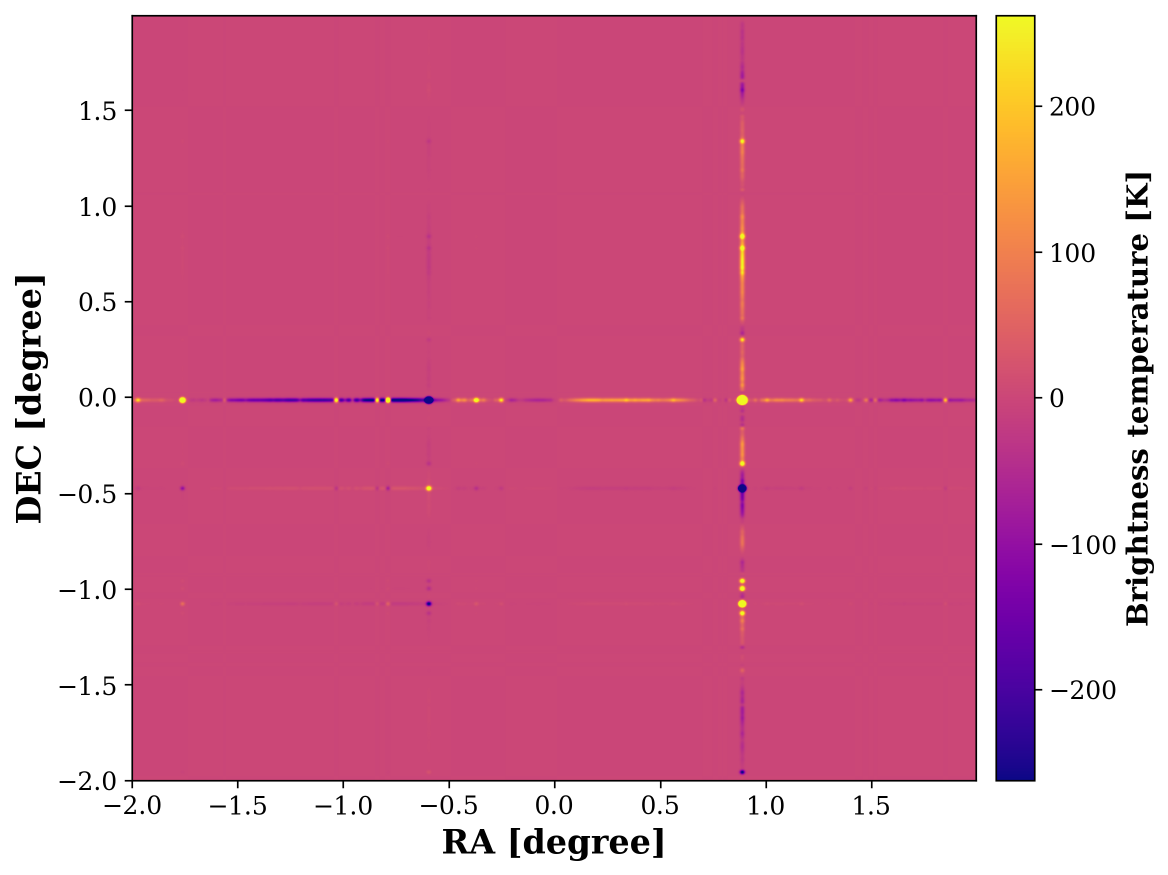}
\end{minipage}
\begin{minipage}{0.4\linewidth}
\centering\includegraphics[width=\textwidth]{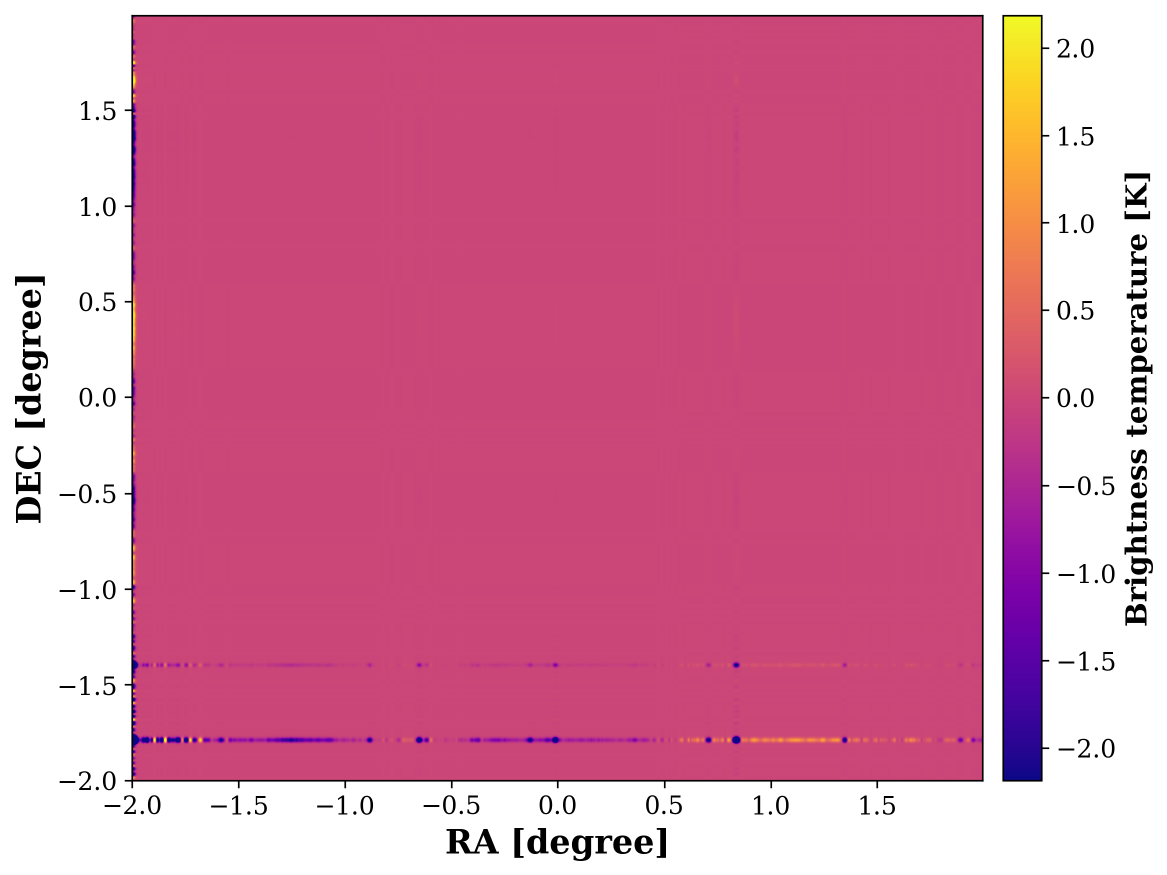}
\end{minipage}
\begin{minipage}{0.4\linewidth}
\centering\includegraphics[width=\textwidth]{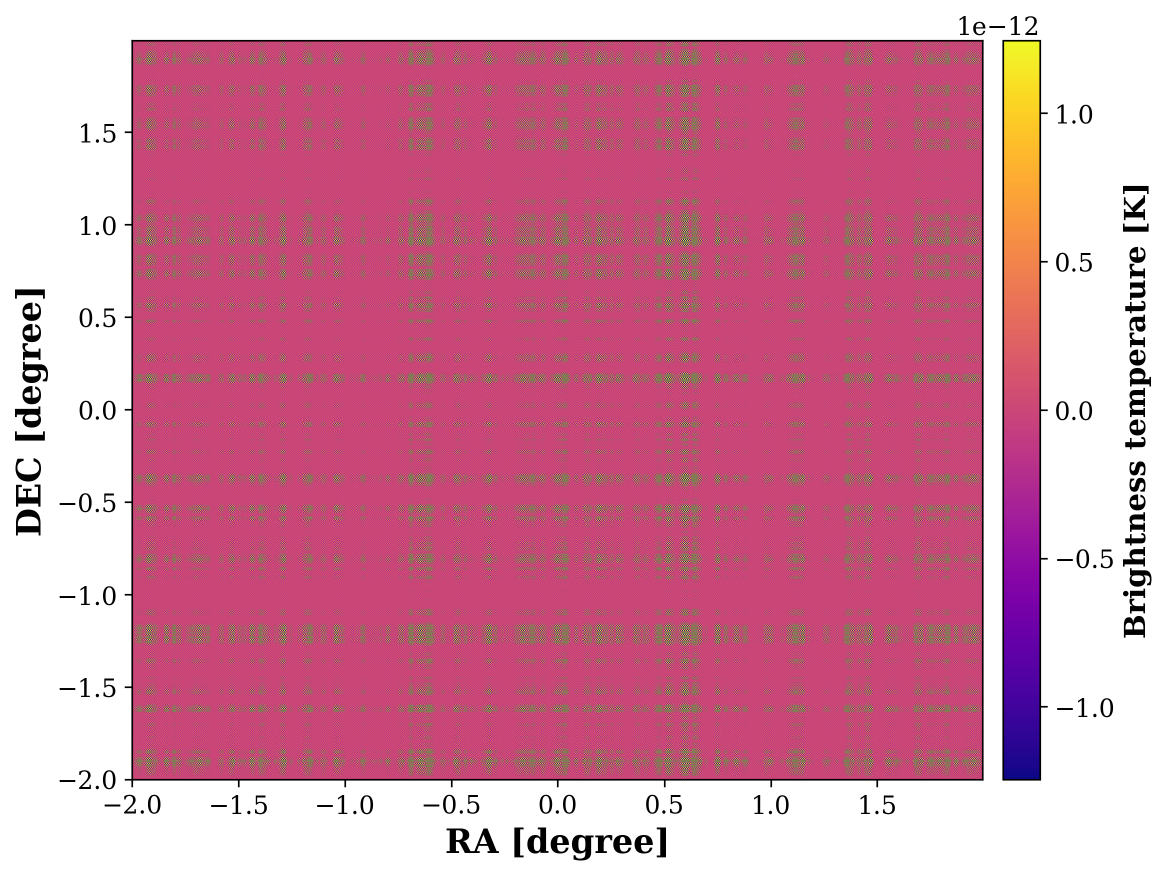}
\end{minipage}
\caption{Visualizing the central frequency slice of the physical components of O-SVD: Rank-1 modes $\mathcal{A}_{(j,j,k)}$ for indices $(1,1,1)$ (top-left), $(2,2,1)$ (top-right), $(1,1,2)$ (bottom-left), and $(900,900,150)$ (bottom-right). The dominant modes (top row) capture high-dynamic-range astrophysical features such as bright point sources and diffuse Galactic emission, while the high-index mode (bottom-right) represents featureless thermal noise. Color scales are adjusted for each panel to emphasize structural details.}
\label{fig:omodes}
\end{figure*}

The physical interpretation of the O-SVD modes is facilitated by the rank-1 decomposition (Eq.~\ref{eq:Ar1}). Here a rank-1 mode with index $(j, j, k)$ is $\mathcal{A}_{(j,j,k)} = s_{jjk} \times_{1} \bm{u}_{kj} \times_{2} \bm{v}_{kj} \times_{3} \bm{u}_{k}$. Figure~\ref{fig:omodes} visualizes the central frequency slice of selected rank-1 components. Dominant modes (e.g., $(1,1,1)$ and $(2,2,1)$) exhibit high-amplitude structures characteristic of bright point sources and diffuse foregrounds, whereas low-amplitude modes (e.g., $(900,900,150)$) appear as featureless noise.

\begin{figure}
\centering
\includegraphics[width=0.45\textwidth]{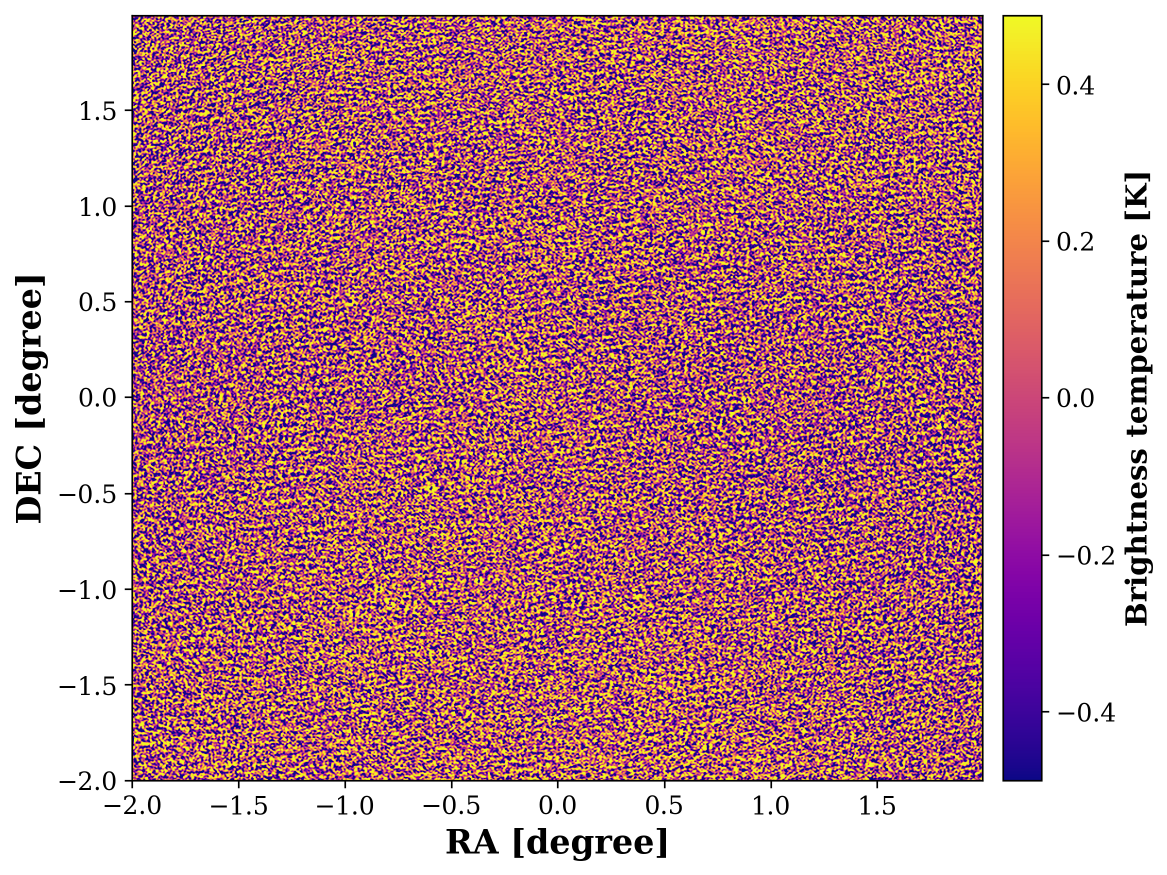}
\caption{The central frequency slice of the residual image cube after O-SVD-based foreground subtraction, with $N_{fg} = 17,652$ modes removed. The residual exhibits the stochastic, Gaussian-like fluctuations characteristic of the cosmological signal and thermal noise.}
\label{fig:rescube}
\end{figure}

The central frequency slice of the residual image cube after removing $N_{fg} = 17,652$ modes is shown in Figure~\ref{fig:rescube}. The resulting cube exhibits the stochastic, noise-like characteristics expected for the recovered 21\,cm signal.

\begin{figure}
\centering
\begin{minipage}{0.9\linewidth}
\centering\includegraphics[width=\textwidth]{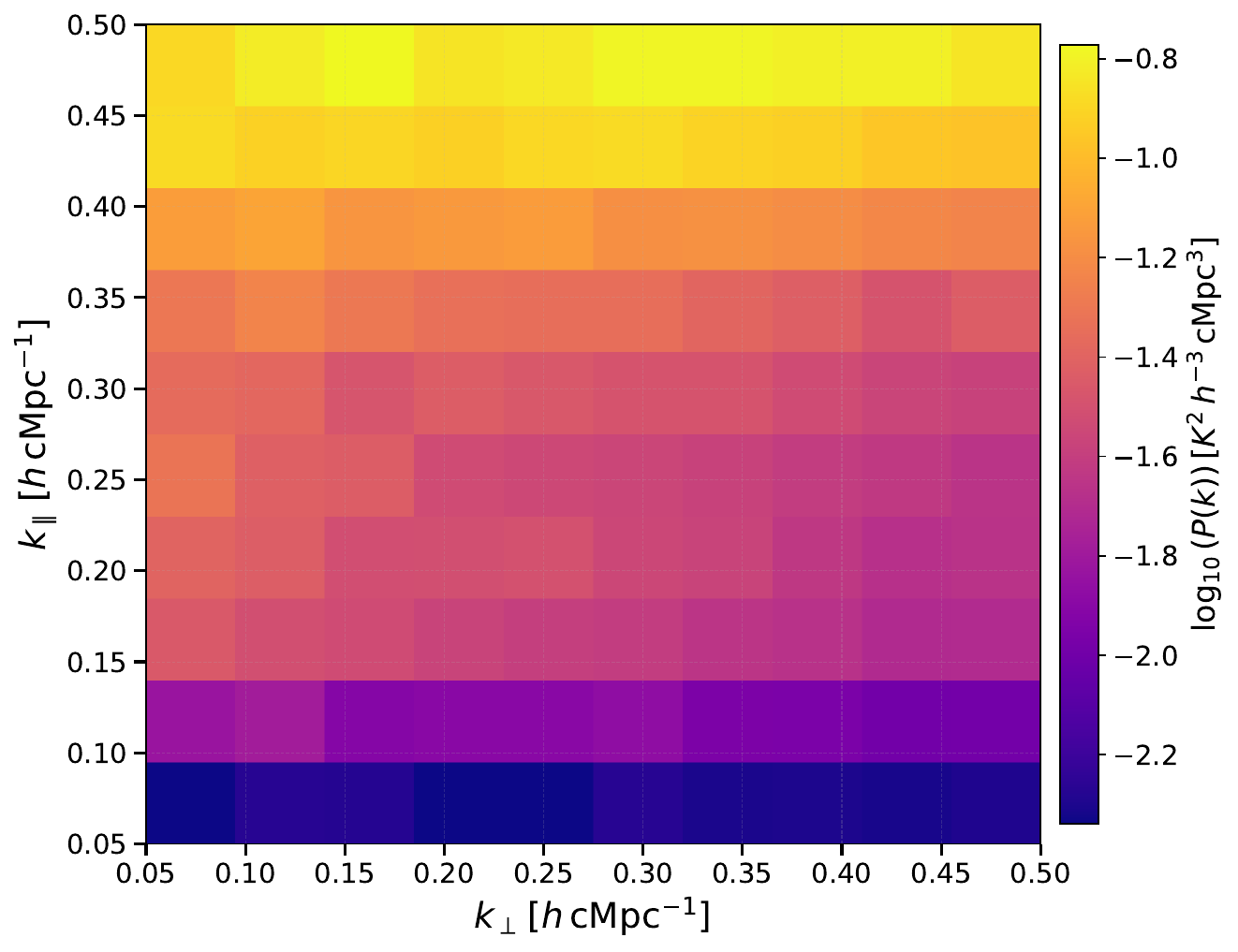}
\end{minipage}\\
\begin{minipage}{0.9\linewidth}
\centering\includegraphics[width=\textwidth]{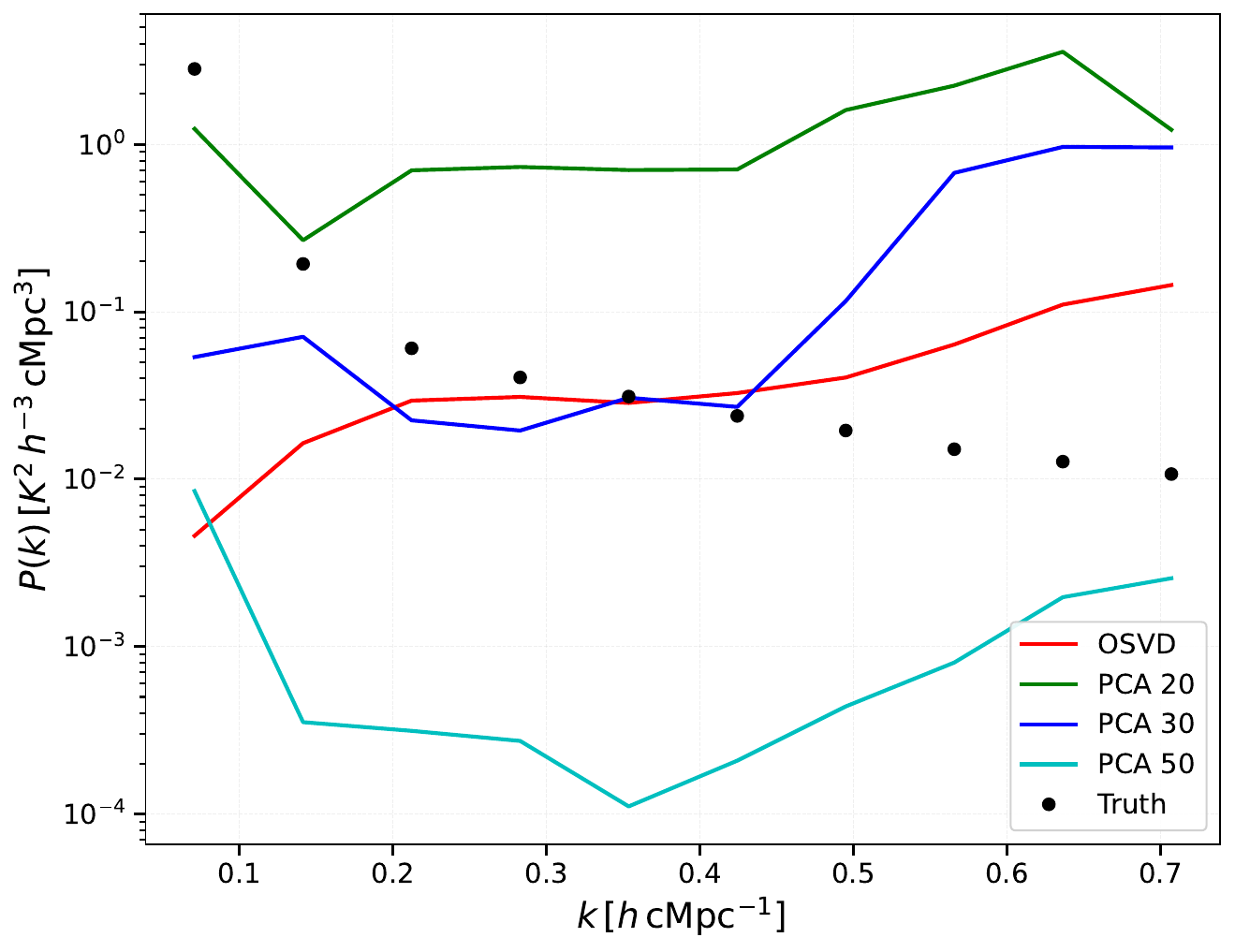}
\end{minipage}
\caption{Top: Cylindrical power spectrum $P(k_\perp, k_\parallel)$ of the O-SVD residual. Bottom: Comparison of the diagonal terms ($k_\parallel = k_\perp$) against the ground-truth EoR power spectrum.}
\label{fig:ps}
\end{figure}

The cylindrical power spectrum $P(k_\perp, k_\parallel)$ estimated from the residual cube is presented in Figure~\ref{fig:ps}. The right panel compares the diagonal terms ($k_\parallel = k_\perp$) of our estimate (red line) against the ground-truth EoR power spectrum. For comparison, the results of the traditional PCA method with 20 modes (green line), 30 modes (blue line), and 50 modes (cyan line) subtracted are also shown. Compared with the traditional PCA method, which shows either high foreground residuals (20 and 30 modes) or significant signal loss (50 modes), our results show excellent agreement with the truth across most $k$-scales. Significant suppression at the largest scales (lowest $k$) is observed, which is a known consequence of signal loss in blind foreground subtraction. This effect can be characterized and corrected using a transfer function approach \citep[e.g.,][]{2013ApJ...763L..20M}. A transfer function $\mathcal{T}(k)$ is defined as the ratio between the reconstructed signal power spectrum and the underlying truth, $\mathcal{T}(k) \equiv P_{\rm clean}(k) / P_{\rm true}(k)$. The estimation of $\mathcal{T}(k)$ relies on a process of mock signal injection. By subjecting the injected mocks to the same foreground cleaning process as the observations, one can use the drop in the measured mock power spectra to estimate the transfer function; see \citet{2023MNRAS.523.2453C} for a step-by-step recipe for constructing and applying an unbiased transfer function.

These results demonstrate the inherent advantages of the O-SVD method over the traditional matrix-based PCA. In the matrix SVD spectrum (Figure~\ref{fig:sigvals}, right), identifying a clear truncation threshold is difficult. We found that even after subtracting 30 PCA modes from the flattened array, the residuals retained significant foreground contamination, as shown in Figure~\ref{fig:ps}. Utilizing the O-SVD relationship (Eq.~\ref{eq:ss}), we find that $N_{fg} = 17,652$ O-SVD modes corresponds approximately to the variance of $N_{fg}^{\rm PCA} \approx 20$ matrix modes. However, the matrix SVD effectively subtracts all O-SVD modes with $k \le 20$ (indicated by the green line in Figure~\ref{fig:sigvals}), including many small variance modes that do not contribute significantly to the foreground power. By contrast, the O-SVD provides a more refined decomposition, allowing for the targeted removal of specific spatial-spectral modes (the region below the black contour). This additional degree of freedom enables O-SVD to isolate foregrounds more effectively while minimizing signal loss in the cosmological signal.

% Our results indicate that the O-SVD method successfully mitigates the dominant foreground components while preserving the underlying 21\,cm signal across a wide range of $k$-modes, demonstrating its robustness against the realistic instrumental systematics present in the SDC3a dataset.

\subsection{Application on Tianlai Data} \label{subsec:tianlai}

\subsubsection{The Tianlai Cylinder Pathfinder Array}
The Tianlai Cylinder Pathfinder Array (TCPA), situated at a radio-quiet site in Xinjiang, China, is a specialized 21\,cm intensity mapping experiment designed to probe the large-scale structure of the Universe during the post-reionization era \citep{2021MNRAS.506.3455W}. The instrument consists of three adjacent, North-South oriented cylindrical reflectors, each 15\,m wide and 40\,m long. The array is equipped with 96 dual-polarization feeds distributed along the focal lines, providing a wide-field survey capability. The TCPA operates within the frequency range of 685--810\,MHz, corresponding to the redshift interval $z \approx 0.77$--$1.03$, with the primary scientific goal of detecting the neutral hydrogen (HI) signal to constrain dark energy through Baryon Acoustic Oscillations (BAO).

\subsubsection{Observations and Data Reduction}
Our analysis is based on a 20-day drift-scan observational dataset acquired by the TCPA in early 2018. The data reduction process, as implemented in the \texttt{tlpipe} pipeline and described in detail by \citet{2021A&C....3400439Z}, consists of the following key stages:
\begin{itemize}
    \item \textbf{RFI Mitigation}: Automated identification and flagging of radio frequency interference (RFI) were performed using a hybrid approach combining the SumThreshold algorithm \citep{2010MNRAS.405..155O} and the scale-invariant rank (SIR) operator \citep{2012A&A...539A..95O}.
    \item \textbf{Calibration}: The complex instrumental gains were calibrated using the bright celestial source Cygnus A as a primary flux calibrator, supplemented by a dedicated noise source for phase stability monitoring \citep{2019AJ....157...34Z}.
    \item \textbf{Map Making}: Sky maps for both XX and YY polarizations were reconstructed using the $m$-mode formalism \citep{2014ApJ...781...57S}. In this work, we focus on the frequency range of 712.9--783.1\,MHz, selected for its relatively high data quality and stability.
\end{itemize}

\subsubsection{Foreground Subtraction in the Angular Power Spectrum Domain}
Rather than operating directly on three-dimensional image cubes, our analysis for the Tianlai dataset is conducted in the angular power spectrum domain using Multi-Frequency Angular Power Spectra (MAPS), denoted as $C_{\ell}(\nu, \nu')$. The MAPS characterizes the cross-frequency correlations of the spherical harmonic coefficients $a_{\ell m}(\nu)$, defined via the relation $\langle a_{\ell m}(\nu) a_{\ell' m'}^*(\nu') \rangle = C_{\ell}(\nu, \nu') \delta_{\ell \ell'} \delta_{mm'}$.

\begin{figure}
\centering
\includegraphics[width=0.45\textwidth]{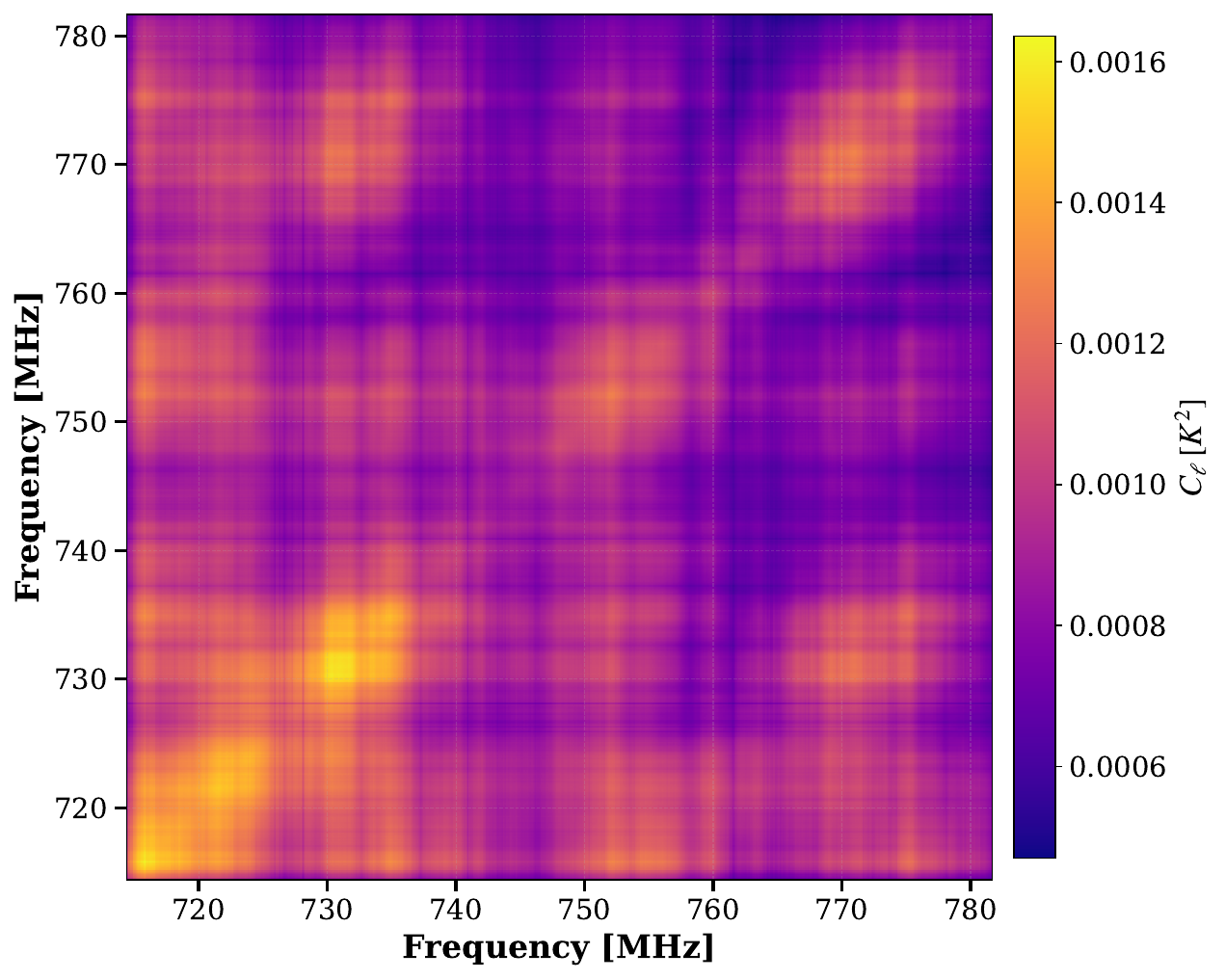}
\caption{Representative frontal slice $C_{\ell=100}(\nu, \nu')$ of the MAPS tensor $\mathcal{C}$ for the Tianlai dataset. The high-amplitude diagonal features and smooth off-diagonal structures are the signature of bright astrophysical foregrounds with long-range frequency coherence.}
\label{fig:cl100}
\end{figure}

In this analysis, we focus exclusively on the YY polarization channel, which exhibited superior spectral smoothness and lower systematic contamination compared to the XX channel in the preliminary data reduction. Given that the cosmological 21\,cm signal is expected to be unpolarized, this selection does not compromise the generality of our findings. We constructed a third-order tensor $\mathcal{C} \in \mathbb{R}^{N_\nu \times N_\nu \times N_\ell}$, where the dimensions correspond to frequency $\nu$, frequency $\nu'$, and multipole $\ell$, respectively. In this representation, each frontal slice $\mathcal{C}(:,:,\ell)$ corresponds to the MAPS matrix $C_{\ell}(\nu, \nu')$ at a specific angular scale. Figure~\ref{fig:cl100} illustrates a representative frontal slice at $\ell=100$, showcasing the dominant foreground characteristics: high-magnitude power concentrated near the diagonal and smooth, long-range frequency correlations. By applying the O-SVD framework to $\mathcal{C}$, the highly correlated foreground components are isolated into the primary multilinear modes, allowing for their effective removal to recover the underlying HI signal and noise residuals.

\begin{figure*}
\centering
\begin{minipage}{0.45\linewidth}
\centering\includegraphics[width=\textwidth]{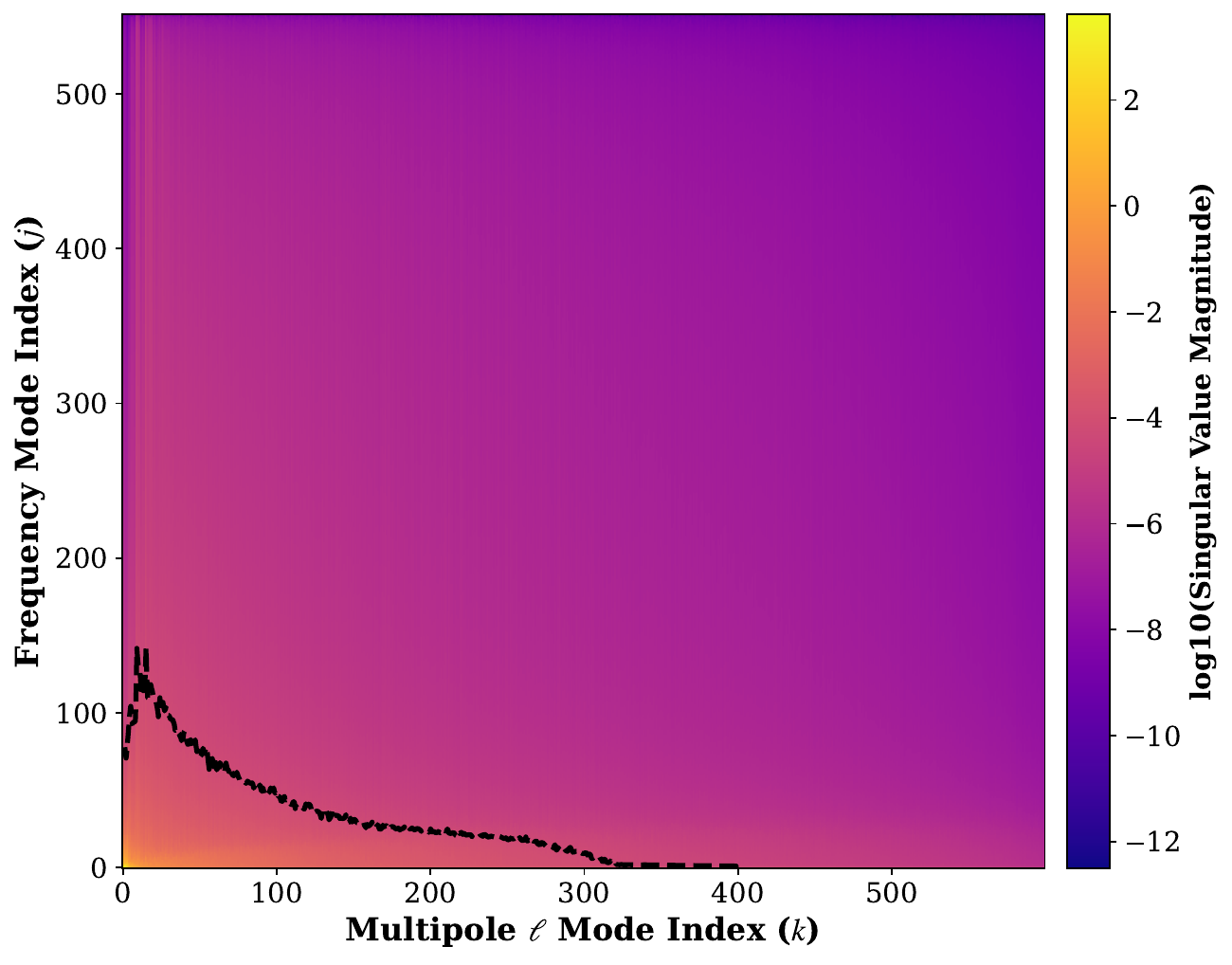}
\end{minipage}
\begin{minipage}{0.45\linewidth}
\centering\includegraphics[width=\textwidth]{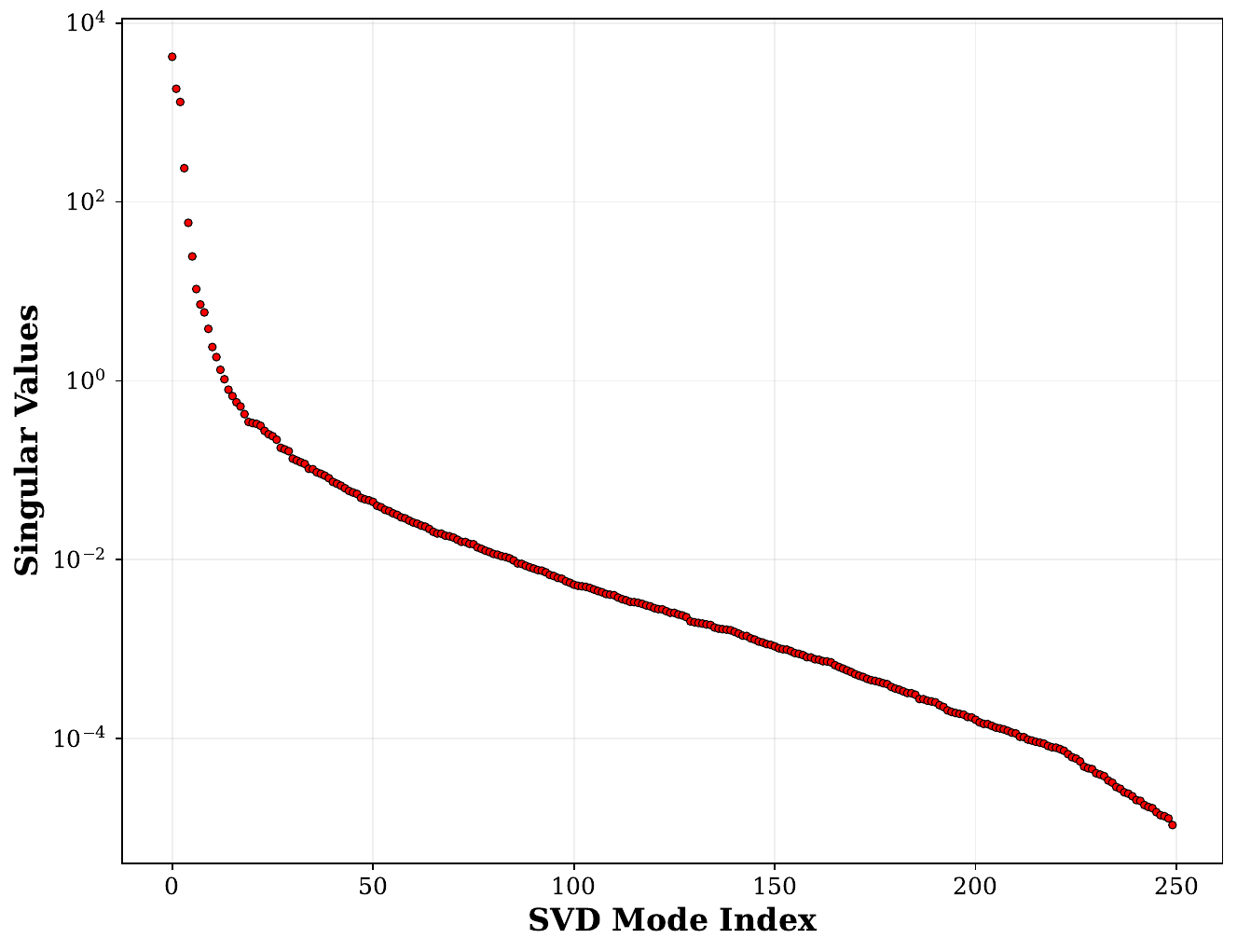}
\end{minipage}
\caption{Comparison of singular value distributions for the Tianlai MAPS tensor: (Left) The structured 2D O-SVD singular value spectrum $s_{jjk}$. (Right) The monotonically decreasing matrix SVD singular values $\sigma_k$, which lacks a clear ``knee'' for optimal threshold selection.}
\label{fig:sigvalscl}
\end{figure*}

The distribution of O-SVD singular values $s_{jjk}$ for the tensor $\mathcal{C}$ is presented in the left panel of Figure~\ref{fig:sigvalscl}, while the right panel shows the singular values $\sigma_{k}$ obtained from a standard matrix SVD of the flattened (mode-3 unfolding) version of the data.

As observed in the matrix SVD distribution (Figure~\ref{fig:sigvalscl}, right), the singular values decrease monotonically without a distinct transition or ``knee'' that would signify an optimal truncation threshold. Conversely, the O-SVD singular value spectrum provides a more structured representation. Adopting the methodology established in Section~\ref{S:dpa}, we utilize the first row of the O-SVD singular values ($s_{11k}$) to identify the noise floor. We identify $k = 400$ as the transition point where the singular values begin to plateau into a noise-like floor. Consequently, we define $s_{11, k=400}$ as the truncation threshold and excise all O-SVD modes with corresponding singular values $s_{jjk} \ge s_{11, k=400}$ (indicated by the region below the black boundary in Figure~\ref{fig:sigvalscl}, left). This process removes 13,283 dominant modes, yielding the foreground-subtracted residual tensor $\mathcal{C}^{\text{res}}$.

\begin{figure*}
\centering
\begin{minipage}{0.45\linewidth}
\centering\includegraphics[width=\textwidth]{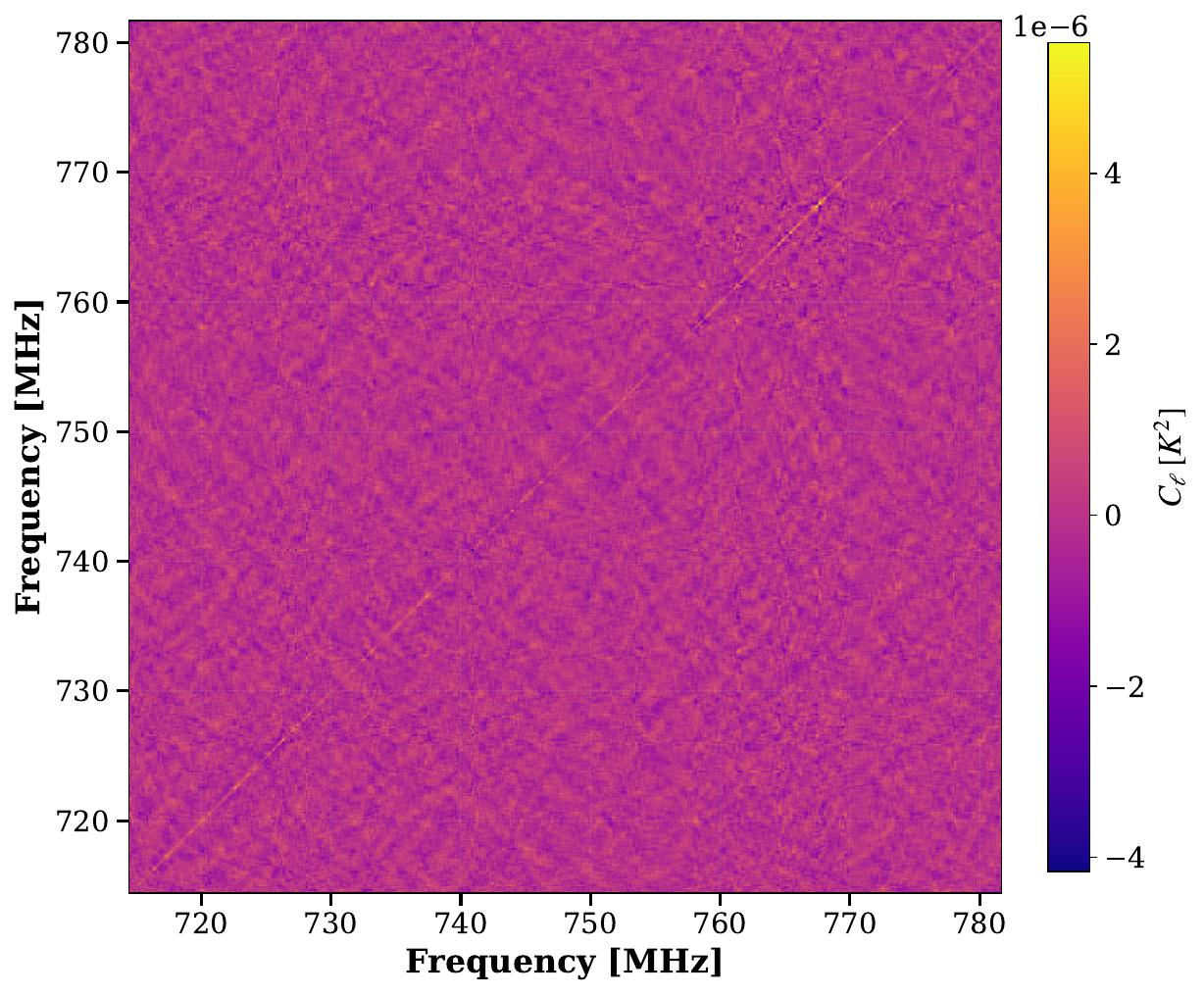}
\end{minipage}
\begin{minipage}{0.45\linewidth}
\centering\includegraphics[width=\textwidth]{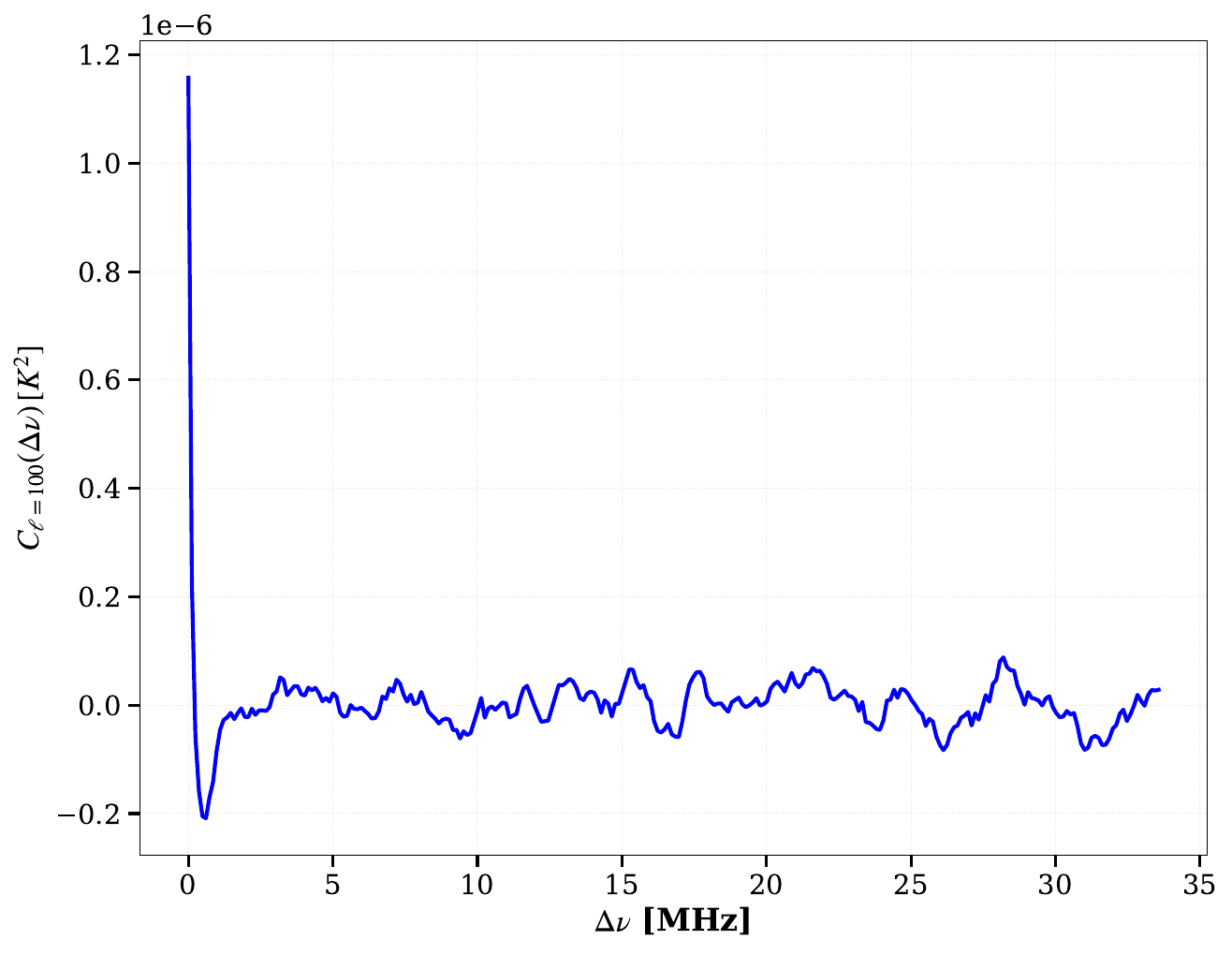}
\end{minipage}
\caption{(Left) Foreground-subtracted residual frontal slice $C^{\text{res}}_{\ell=100}(\nu, \nu')$. (Right) Averaged residual power as a function of frequency separation $\Delta \nu$, demonstrating the rapid decoherence of the signal compared to the original foreground-dominated data.}
\label{fig:res100}
\end{figure*}

The efficacy of the O-SVD subtraction is demonstrated in Figure~\ref{fig:res100} (left), which displays the residual frontal slice $C^{\text{res}}_{\ell=100}(\nu, \nu')$. The residual map shows a marked concentration of power along the diagonal, with significantly attenuated off-diagonal features, consistent with the expected behavior of the 21\,cm signal and thermal noise. While ideal noise is frequency-independent and strictly diagonal, instrumental systematics and pipeline artifacts may introduce residual correlations. The frequency correlation length $\Delta \nu$ of the 21\,cm signal is intrinsically linked to the angular scale; at $\ell \sim 100$, the signal typically decoheres beyond $\Delta \nu \sim 1$\,MHz, whereas at $\ell \sim 10^{3}$, this occurs at $\sim 0.1$\,MHz \citep{2005MNRAS.356.1519B,2007MNRAS.378..119D}. This behavior is further quantified in Figure~\ref{fig:res100} (right), which plots the averaged residual power $C^{\text{res}}_{\ell=100}(\Delta \nu)$ as a function of frequency separation $\Delta \nu = |\nu_{j} - \nu_{i}|$.

The angular power spectrum $C_{\ell}(\Delta \nu)$ is related to the cylindrical 2D power spectrum $P(k_{\perp}, k_{\parallel})$ via the flat-sky approximation \citep{2005MNRAS.356.1519B,2007MNRAS.378..119D,2018MNRAS.474.1390M}:
\begin{equation} \label{eq:pkk}
  P(k_{\perp}, k_{\parallel}) = r_{c}^2 r'_{c} \int d(\Delta \nu) e^{-i k_\parallel r_{c}' \Delta \nu} C_\ell(\Delta \nu),
\end{equation}
where $k_{\perp} = \ell/r_{c}$, $r_{c}$ is the comoving distance to the center of the light-cone, and $r'_{c} = \frac{dr}{d\nu} |_{r_{c}}$. Although Equation~\ref{eq:pkk} allows for the computation of $P(k_{\perp}, k_{\parallel})$, we refrain from presenting the resulting power spectra here. Given the limited duration of the early TCPA dataset (20 days), the signal-to-noise ratio is insufficient for a definitive 21\,cm detection; the residuals remain dominated by thermal noise and low-level systematic residuals.

Nevertheless, these results successfully demonstrate that O-SVD can be effectively applied to high-dimensional datasets represented as tensors beyond the standard image-cube format. This highlights the flexibility of the O-SVD framework in addressing foreground mitigation across diverse observational representations in 21\,cm cosmology.

\section{Discussion} \label{sec:dis}
In this study, we have demonstrated that the O-SVD framework provides a robust and mathematically rigorous approach to foreground subtraction, offering enhanced flexibility compared to conventional matrix-based SVD methods. In contemporary 21\,cm cosmology, Principal Component Analysis (PCA) remains a cornerstone of foreground mitigation pipelines due to its computational efficiency and simplicity as a blind signal separation (BSS) technique. For instance, in the SKA SDC3a challenge, several leading pipelines—including Foregrounds-FRIENDS, HIMALAYA, and REACTOR—incorporated PCA as a critical processing stage, while others like HAMSTER utilized it for diagnostic and multi-stage analysis \citep{2025MNRAS.543.1092B}.

The emerging synergy between traditional blind subtraction and machine learning further underscores the importance of efficient dimensionality reduction. In many hybrid frameworks, PCA is employed as a vital pre-processing step to compress the vast dynamic range between astrophysical foregrounds and the faint cosmological signal, which is often a prerequisite for the effective training of deep neural networks. Notable examples include the \texttt{deep21} method \citep{2021JCAP...04..081M}, which utilizes a U-Net architecture to reconstruct 21\,cm maps from PCA-reduced inputs, and recent work by \citet{2022ApJ...934...83N} showing that U-Net-based recovery is significantly enhanced when operating on PCA-subtracted residuals, particularly in the presence of complex, beam-induced chromatic systematics.

The intrinsic mathematical link between the O-SVD of a third-order oriented tensor and the matrix SVD of its mode-3 unfolding suggests that O-SVD can serve as a seamless, ``plug-and-play'' replacement for conventional PCA. By leveraging additional spatial-spectral filtering degrees of freedom, O-SVD allows for a more precise isolation of foreground modes while minimizing signal loss. We anticipate that integrating O-SVD into existing pipelines—including those serving as front-ends for deep learning models—will yield superior signal-foreground separation without requiring fundamental changes to the underlying data processing architecture.

To quantify the computational cost, we provide a detailed complexity analysis for both SVD and O-SVD. For a data cube of size $N_x \times N_y \times N_\nu$:

\begin{itemize}
    \item \textbf{Traditional SVD (PCA)}: The 3D cube is unfolded to a $N_\nu \times (N_x N_y)$ matrix. SVD of an $m \times n$ matrix takes $O(m^2 n + m^3)$ when $m \le n$, giving:
    \begin{itemize}
        \item Complexity: $O(N_\nu^2 N_x N_y + N_\nu^3)$
    \end{itemize}
    \item \textbf{O-SVD}: Two-stage process:
    \begin{itemize}
        \item Stage 1: Same as traditional SVD: $O(N_\nu^2 N_x N_y + N_\nu^3)$
        \item Stage 2: Each spatial mode is reshaped to an $N_x \times N_y$ matrix and decomposed via SVD: $O(N_\nu \cdot (N_x^2 N_y + N_x^3))$ assuming $N_x \le N_y$
        \item Total complexity: $O(N_\nu^2 N_x N_y + N_\nu^3 + N_\nu \cdot (N_x^2 N_y + N_x^3))$
    \end{itemize}
\end{itemize}

For our SDC3a dataset ($900 \times 900 \times 150$), the additional computational cost of a full O-SVD decomposition is approximately 13$\times$. In practice, since usually $N_\nu \ll (N_x N_y)$ and only the largest singular values need to be computed in the second stage, the computational complexity can be reduced to $O(N_\nu^2 N_x N_y)$ for traditional SVD versus $O(N_\nu^2 N_x N_y + N_\nu \min(N_x, N_y)^3)$ for O-SVD, which is 6--7$\times$ for the SDC3a dataset. This trade-off between computational cost and recovery quality is acceptable for our applications, as the modest increase in computation time is justified by the improved foreground separation performance. While the computational complexity of O-SVD is higher than that of standard matrix operations, recent advancements in randomized algorithms for oriented tensors \citep[e.g.,][]{Ding2022ARS} have shown that this overhead can be substantially reduced through stochastic approximations while maintaining rigorous error bounds. Such techniques make O-SVD increasingly viable for the massive datasets expected from the next generation of radio interferometers like the SKA.

Beyond O-SVD, the field of multilinear algebra offers a diverse array of tensor decomposition methods—such as CANDECOMP/PARAFAC (CP) \citep{1970Analysis,harshman70}, Higher-Order SVD (HOSVD) \citep{Lathauwer2000AMS,1966Some}, and Tensor-Train (TT) \citep{doi:10.1137/090752286}—that could provide even more compact and physically motivated representations of astronomical data. While historically limited by conceptual complexity and computational demands, these methods represent a promising frontier for 21\,cm foreground mitigation and broader astrophysical signal processing. We hope this study serves as a catalyst for the further exploration of advanced multilinear methodologies within the astronomical community.

\section{Summary and Conclusions} \label{sec:sum}
In this paper, we have introduced the Oriented Singular Value Decomposition (O-SVD) as a robust and mathematically rigorous framework for foreground mitigation in 21\,cm intensity mapping. By preserving the inherent multilinear structure of astronomical datasets, O-SVD provides a theoretically grounded alternative to traditional matrix-based decomposition methods that rely on data flattening. To the best of our knowledge, this work represents the first systematic application of such tensor-based decomposition techniques to the challenge of separating the faint cosmological 21\,cm signal from dominant astrophysical emissions.

The efficacy and versatility of the O-SVD framework have been demonstrated through its successful application to both high-fidelity simulations and real observational data. In the context of the SKA SDC3a dataset, the method proved capable of isolating the Epoch of Reionization (EoR) signal in the presence of complex instrumental systematics and high-dynamic-range foregrounds. Furthermore, our application to the Tianlai Cylinder Pathfinder Array observational data confirms that O-SVD can be effectively extended to the angular power spectrum domain, successfully removing smooth-spectrum foregrounds from cross-frequency correlations.

The O-SVD framework offers a unified and flexible architecture capable of processing diverse data representations, ranging from spatial-frequency image cubes to multi-frequency angular power spectra. Future research will explore the integration of automated mode truncation strategies based on objective information-theoretic criteria—such as the Akaike Information Criterion (AIC), Bayesian Information Criterion (BIC), or Generalized Cross-Validation (GCV)—to further minimize signal bias and enhance the reproducibility of foreground subtraction. We anticipate that this work will serve as a catalyst for the broader adoption of advanced multilinear algebra and tensor-based methodologies within the astronomical community, providing powerful new tools for the era of precision 21\,cm cosmology.

\begin{acknowledgments}
  We acknowledge the support by the National Natural Science Foundation of China (Nos. 12303004, 12203061, 12361141814 and 12273070), National SKA Program of China (Nos. 2022SKA0110100, 2022SKA0110101, 2020SKA0110401), and the Chinese Academy of Sciences ZDKYYQ20200008.
\end{acknowledgments}

\software{astropy \citep{2013A&A...558A..33A},  
          tlpipe \citep{2021A&C....3400439Z},
          tools21cm \citep{2020JOSS....5.2363G},
          radio\_beam,
          OSKAR \citep{2009wska.confE..31D},
          WSClean \citep{2014MNRAS.444..606O},
          21cmFAST \citep{2011MNRAS.411..955M,2020JOSS....5.2582M}
          }

\clearpage
\appendix

\section{Algorithm for O-SVD} \label{S:alg}

The O-SVD algorithm for a third-order tensor is summarized in Algorithm~\ref{alg:osvd}.
% This decomposition captures the spatial correlations within each spectral mode (or slice) after transforming the tensor into its principal spectral components.

\begin{algorithm}[!ht]
\caption{O-SVD}
\label{alg:osvd}
\KwIn{Third-order tensor $\mathcal{A} \in \mathbb{C}^{I_1 \times I_2 \times I_3}$}
\KwOut{Unitary matrix $\bm{U}^{(3)} \in \mathbb{C}^{I_3 \times I_3}$, and tensors $\mathcal{U}^{(3)} \in \mathbb{C}^{I_1 \times I_1 \times I_3}$, $\mathcal{S}^{(3)} \in \mathbb{C}^{I_1 \times I_2 \times I_3}$, $\mathcal{V}^{(3)} \in \mathbb{C}^{I_2 \times I_2 \times I_3}$}
$[\bm{U}^{(3)}, \bm{\Sigma}^{(3)}, \bm{V}^{(3)}] = \text{svd}[\bm{A}_{(3)}]$ \tcp*[r]{Mode-3 unfolding and SVD}
$R_3 = \text{rank}(\bm{\Sigma}^{(3)})$\;
$\hat{\mathcal{A}} = \mathcal{A} \times_3 \bm{U}^{(3)H}$ \tcp*[r]{Transform to principal components}
\For{$k = 1, \dots, R_3$}{
    $[\bm{U}, \bm{S}, \bm{V}] = \text{svd}[\hat{\mathcal{A}}(:,:,k)]$ \tcp*[r]{SVD of each spatial slice}
    $\mathcal{U}^{(3)}(:,:,k) = \bm{U}$\;
    $\mathcal{S}^{(3)}(:,:,k) = \bm{S}$\;
    $\mathcal{V}^{(3)}(:,:,k) = \bm{V}^H$\;
}
\For{$k = R_3 + 1, \dots, I_3$}{
    $\mathcal{U}^{(3)}(:,:,k) = \bm{0}$, $\mathcal{S}^{(3)}(:,:,k) = \bm{0}$, $\mathcal{V}^{(3)}(:,:,k) = \bm{0}$ \tcp*[r]{Zero-padding for remaining slices}
}
\end{algorithm}

\bibliography{osvd}{}
\bibliographystyle{aasjournalv7}

\end{document}